\documentclass{acmart} % review

\usepackage[T1]{fontenc}
\usepackage[utf8]{inputenc}
\usepackage{booktabs}
\usepackage[english]{babel}%texlive-lang-german
\usepackage{listings}
\usepackage{verbatim}

\pdfoutput=1

\newcommand{\bbN}{\mathbb{N}}
\newcommand{\bbZ}{\mathbb{Z}}

\newcommand{\mbB}{\mathbf{B}}

\newcommand{\mbU}{\mathbf{U}}
\newcommand{\mbV}{\mathbf{V}}

\newcommand{\mubA}{\underline{\mathbf{A}}}
\newcommand{\mubB}{\underline{\mathbf{B}}}
\newcommand{\mubC}{\underline{\mathbf{C}}}

\newcommand{\mcI}{\mathcal{I}}
\newcommand{\mcJ}{\mathcal{J}}

\newcommand{\mcW}{\mathcal{W}}

\newcommand{\mbb}{\mathbf{b}}
\newcommand{\mbi}{\mathbf{i}}

\newcommand{\mbn}{\mathbf{n}}
\newcommand{\mbw}{\mathbf{w}}
\newcommand{\mbv}{\mathbf{v}}
\newcommand{\mbu}{\mathbf{u}}

\newcommand{\mbo}{\mathbf{o}}
\newcommand{\mbf}{\mathbf{f}}

\newcommand{\mbpi}{\boldsymbol{\pi}}
\newcommand{\mbpsi}{\boldsymbol{\psi}}

\newcommand{\mbtau}{\boldsymbol{\tau}}

\newcommand{\mbvarphi}{\boldsymbol{\varphi}}

\newcommand{\tsmall}[1]{\texttt{\small {#1}}}

\setcopyright{none}

\lstdefinelanguage{cplusplus}
{ % \small, \scriptsize \footnotesize 
	language=C++, %
	basicstyle=\small\ttfamily,% \singlespacing
	stringstyle=\color{white}\small\ttfamily,
	numbersep=5pt,
	keepspaces=true,
	tabsize=2,
	breaklines=false,
	breakatwhitespace=false,
	title=\lstname,
	numberstyle=\tiny\color{black},
	commentstyle=\small\ttfamily, % \color{Snow4}
	stringstyle=\small\ttfamily, % \color{Purple4}
	morecomment=[l]{//},
	morecomment=[s]{/*}{*/},
	morecomment=[n]{(*}{*)},
	alsoletter={.}
	sensitive=true,
	classoffset=0,
	morekeywords={constexpr,decltype},
	keywordstyle=\bfseries, %\color{Yellow4}, %\bfseries%\color{Yellow4}, \color{black}\footnotesize \color{Blue4}
	classoffset=1,
	keywordstyle=\bfseries, %\color{Blue4}, %\color{Magenta4},\bfseries \color{Green4}
	morekeywords={value_type, reference, tensor, tensor_view, tensor_view_t, range, domain, ptrdiff_t, vector, allocator, std, fhg, plus, cout, endl, complex,iterator, const_iterator, difference_type, size_type, size_t, const_reference, pointer, const_pointer, shape, offset, layout, multi_iterator}, % ,std,
	classoffset=0,
	showtabs=false,
	showspaces=false,
	showstringspaces=false,  
	escapeinside={\%*}{*)},
	belowskip=-5pt, 
	abovecaptionskip=0pt, 
	belowcaptionskip=0pt,
	columns=fullflexible,
	captionpos=b
}

\acmJournal{TOMS}
\acmVolume{0}
\acmNumber{0}
\acmArticle{0}
\acmYear{2017}
\acmMonth{0}
\copyrightyear{2017}

\begin{document}

\lstset{language=cplusplus}
\citestyle{acmauthoryear}

%\title[A Flexible C++ Tensor Framework]{TLib: A Flexible C++ Tensor Framework for Fast Algorithm Prototyping}
\title[TLib: A Flexible C++ Tensor Framework]{TLib: A Flexible C++ Tensor Framework for Numerical Tensor Calculus}    

\author{Cem Bassoy}
\affiliation{%
  \institution{Fraunhofer Institute of Optronics, System Technologies and Image Exploitation}
  %\streetaddress{Gutleuthausstraße 1}
  \city{Ettlingen}
  %\state{VA}
  %\postcode{76275}
  \country{Germany}
}

%The representation of tensors (in particular, with not too large storage requirements) is one goal of the efficient numerical treatment of tensors. Another goal is the efficient performance of tensor operations. In the case of matrices, we apply matrix-vector and matrix-matrix multiplications and matrix inversions. The same operations occur for tensors, when the matrix is given by a Kronecker matrix and the vector by a tensor. Besides of these operations there are entry-wise multiplications, convolutions etc.

%tensors are treated as multidimensional arrays or n-way arrays with 

\begin{abstract}
	Numerical tensor calculus comprise basic tensor operations such as the entrywise addition and contraction of higher-order tensors.
	%using \tsmall{C++}
	%
	We present, TLib, flexible tensor framework with generic tensor functions and tensor classes that assists users to implement generic and flexible tensor algorithms in C++.  
	The number of dimensions, the extents of the dimensions of the tensors and the contraction modes of the tensor operations can be runtime variable.
	Our framework provides tensor classes that simplify the management of multidimensional data and utilization of tensor operations using object-oriented and generic programming techniques.
	Additional stream classes help the user to verify and compare of numerical results with MATLAB. 
	Tensor operations are implemented with generic tensor functions and in terms of multidimensional iterator types only, decoupling data storage representation and computation.
	The user can combine tensor functions with different tensor types and extend the framework without further modification of the classes or functions. 
	We discuss the design and implementation of the framework and demonstrate its usage with examples that have been discussed in the literature.
\end{abstract}


%
% End generated code
%

%\keywords{TODO}

\maketitle

\section{Introduction}
%\structure{What is a multidimensional array?}

In modern mathematics a higher-order tensor is defined as an element of a tensor product space~\cite{lim:2017:hypermatrices,silva:2017:tensors}. Higher-order tensors are coordinate-free in an abstract fashion without choosing a basis of the tensor product space. In the realm of numerical tensor calculus, higher-order tensors with a coordinate representation are considered~\cite{lim:2017:hypermatrices, hackbusch:2014:numerical.tensor.calculus}. The bases are chosen implicitly, and the values of some measurements are then recorded in the form of a multidimensional array. We define a multidimensional array as entity that holds a set of data all of the same type whose elements are arranged in a rectangular pattern. In some cases higher-order tensors are referred to as hypermatrices, $N$-way arrays or $N$-dimensional table of values~\cite{lim:2017:hypermatrices, cichocki:2009:tensor, lathauwer:2000:multilinearsvd} where $N$ is order, i.e. the number of dimensions.  

Basic tensor operations are the tensor-tensor, tensor-matrix, tensor-vector multiplication, the inner and outer product of two tensors, the Kronecker, Hadamard and Khatri-Rao product~\cite{cichocki:2009:tensor,lim:2017:hypermatrices}. Common methods utilizing tensor operations are e.g. the higher order decompositions or to calculate the eigenvalues or singular values of a higher-order tensor~\cite{lathauwer:2000:multilinearsvd, kolda:2009:decompositions, cui:2014:allrealeigenvalues, ng:2009:finding.largest.eigenvalue}. Other types of tensor decomposition are the CP- (Canonical-Decomposition/Parallel-Factor-Analysis)~\cite{harshman:1994:parafac, faber:2003:recent} and Tucker-Decomposition~\cite{tucker:1966:notes, kim:2007:nonnegative} which are mainly used within the field of psychometrics and chemometrics. Other areas of application are signal processing~\cite{savas:2007:handwritten, fitzgerald:2005:tensorfactorization}, computer graphics~\cite{valisescu:2002:facialrecognition, suter:2013:volumevisualization} and data mining~\cite{kolda:2008:datamining, rendle:2009:datamining}. 

Many general-purpose programming languages such as \tsmall{C}, \tsmall{C++} or \tsmall{Fortran} support multidimensional arrays as built-in data structures with which elements are accessed in a convenient manner. Yet, built-in data types might not meet the requirements of an application. For instance, the \tsmall{C++} built-in multidimensional array is not a good fit if the application requires the number of dimensions to be runtime variable. A very common approach is to provide a library that extends the general-purpose language with user-defined data types and functions. In most cases, the interface of the libraries are designed to be close to the notation that is used within a field of application~\cite{Mernik:2005:DSL}. Usually, \tsmall{C++} is chosen to be the host language that is extended with user-defined data types. The key feature of \tsmall{C++} is that it enables the programmer to apply object-oriented and generic programming techniques with a simple, direct mapping to hardware and zero-overhead abstraction mechanisms \cite{stroustrup:2012:foundations,gregor:2006:concepts}. Functions and types can be parametrized in terms of types and/or values supporting parametric polymorphism and allow software to be general, flexible and efficient~\cite{stroustrup:2012:software.development}.

We would first like to introduce \tsmall{C++} libraries that are related to our framework and depict similar high-level interfaces. POOMA, described in~\cite{reynders:1998:pooma}, is perhaps one the first \tsmall{C++} frameworks that have been designed to support arrays with multiple dimensions including tensor operations. Multidimensional arrays are generic data types where the number of dimensions are compile-time parameters. The framework supports high-level expressions for first-level tensor operations. The library described in~\cite{landry:2003:tensorlibrary} offers a tensor class that is designed to support classical applications found in quantum mechanics. Tensor functions for high-level tensor operations are provided. However, the framework only support tensors up to four dimensions with four elements in each dimension limiting the application of the framework to classical applications of numerical tensor calculus. In~\cite{garcia:2005:multiarray}, the design of generic data types for multidimensional arrays is discussed, including the addressing elements and subdomains (views) of multidimensional arrays with first- and last-order storage formats. The order and data type are compile-time (template) parameters. They offer iterators for their multidimensional arrays and suggest to parametrize their tensor algorithms in terms of multidimensional array types that generate stride-based iterators for the current dimension of the recursion. In~\cite{andres:2010:runtime} an implementation of a multidimensional array and iterators are presented where the order and the dimension extents of tensor types are runtime parameters. The paper also discusses addressing functions, yet for the first- and last-order storage format only. Please note that~\cite{reynders:1998:pooma, garcia:2005:multiarray, andres:2010:runtime} do not support higher-order tensor operations for numerical tensor calculus.

Most \tsmall{C++} frameworks supporting tensor contraction provide a convenient notation that is close to Einstein's summation convention. The Cyclops-Tensor-Framework (CT) described in~\cite{solomonik:2013:cyclops} offers a library primarily targeted at quantum chemistry applications~\cite{solomonik:2013:cyclops}. The order and the dimensions of their tensor data structures are dynamically configurable, while the interface omits the possibility to set a data layout or a user-defined data type. The tensor contractions are performed by index foldings with matrix operations where the indices are adjusted with respect to the tensor operation. The interface for specifying the contraction is similar to the one provided by the MiaArray library (LibMia) discussed in~\cite{harrison:2016:numeric_tensor_framework} using either strings (CT) or objects (LibMia, Blitz). For instance, after having instantiated tensor objects, the $2$-mode multiplication of a three dimensional with a matrix is given by \tsmall{C["ijk"]=A["ilj"]*B["kl"]} in case of the CT framework, \tsmall{C(i,j,k)=A(i,l,k)*B(k,l)} in case of the LibMia framework, or \tsmall{C(i,j,k)=sum(A(i,l,j)*B(k,l),l)} in case of the Blitz framework~\cite{veldhuizen:1998:arrays} where the type of the variables \tsmall{i,j,k,l} of the example code snippets are user-defined. The notation necessitates the order of the arrays to be known before compile-time and does not allow the contraction mode to be runtime-variable. The implementation of tensor algorithms such as the higher-order singular value decomposition for instance, performs a sequence of $k$-mode tensor-times-vector multiplications where the mode $k$ depends on an induction variable~\cite{lathauwer:2000:multilinearsvd, bader:2006:algorithm862}.

Our framework fills this need with flexible and generic tensor classes and functions where the order, extents of the dimensions and the contraction mode(s) can be runtime variable. The interfaces of the generic tensor functions are similar to the ones provided by the \tsmall{Matlab} library that is presented in~\cite{bader:2006:algorithm862}. The toolbox provides tensor classes and tensor operations for prototyping tensor algorithms in \tsmall{Matlab}. The execution of some higher-order tensor operations such as tensor-tensor-multiplication requires a tensor to be converted or unfolded into matrix. The unfolding is performed with respect to the mode of tensor operation and requires additional memory space for the unfolded tensor. However, once the unfolding is accomplished, fast matrix multiplications can be used to perform the tensor contraction. This approach is also applied in frameworks such as in~\cite{dinapoli:2014:towards.efficient.use}. The generic tensor functions of our framework execute tensor operations in-place for tensor types including domains of tensors without the process of unfolding. They also support a set of storage formats including the first- and last-order storage formats. Please note that we do not intend to replace any of the previously mentioned works, but to provide a flexible and extensible library that allows to easily validate numerical results with the Matlab toolbox provided by~\cite{bader:2006:algorithm862}. 

We would like to limit our discussion to the software design of a framework for basic tensor operations. The design and implementation of high-performance algorithms exploiting data locality and the parallel processing capabilities of multi-core processors lies beyond the scope of this paper. We would like to refer to~\cite{springer:2016:design.gett} in which a method for fast tensor contractions with arbitrary order and dimensions is described. The optimization techniques are similar to the ones applied for matrix-matrix multiplication. The paper presents a TCCG, a \tsmall{C++} code generator that generates optimized \tsmall{C++}  code for the tensor multiplication. The code generation requires an input file where the contracting dimensions and mode of the multiplication, order, dimensions of the multidimensional arrays are specified. The work presented in~\cite{li:2015:adaptive.inplace.tensor.times.matrix.multiply} follows a similar approach and provides code generator that is called InTensLi.

Our framework consists of a software stack with two main components. The upper part of the software stack contains flexible tensor classes with runtime variable number of dimensions and dimension extents. The classes and their member functions simplify the management of multidimensional data and the selection of single elements or multidimensional domains using generic and object-oriented programming techniques. The data layout of the multidimensional data can be adjusted at runtime as well, supporting a class of storage layouts include the first- and last-order storage formats. Member functions of the tensor classes encapsulate generic tensor functions of the lower software stack components and help the programmer to write tensor algorithms independent of the storage layout of the tensors. We have used ad-hoc polymorphism, i.e. operator overloading, for lower-level (entrywise) tensor operations in order to enable the user to write tensor algorithms close to the mathematical notation. Member functions encapsulating higher-level tensor operations have an interface that is similar to the one provided by the toolbox discussed in~\cite{bader:2006:algorithm862}. The upper component also provides stream classes with which numerical results can be validated in \tsmall{Matlab}. The lower part of the software stack includes generic tensor function that are defined in terms of multidimensional iterator types only. The decoupling of data storage representation and computation with multidimensional iterators allows different tensor types to be handled with same tensor function. Our framework provides its own multidimensional iterator that is one possible interface between the tensor classes and generic tensor functions. The user can extend the framework with tensor functions without modifying the tensor template classes and provide his own tensor and iterator types. In summary, the main contributions of our work are as follows:
\begin{itemize}
	\item %Higher-level tensor functions are implemented with runtime-variable modes in a recursive fashion with similar interfaces described in~\cite{bader:2006:algorithm862} where the mode of the tensor multiplication and the order, dimensions, storage format, index offsets of the tensor can therefore be runtime parameters. 
	Our implementation of higher-level tensor operations allows the mode of the tensor multiplication and the order, dimensions, storage format, index offsets of the tensor to be runtime parameters. The user can implement tensor algorithms with arguments that depend on runtime-variable parameters.	
	\item %All tensor template functions are parametrized in terms of multidimensional iterator types, decoupling storage representation of data and computation. The user can provide its own tensor or iterator types with a variety data layouts including tensor references that are processed by the same tensor function eliminating the need to define multiple algorithms for similar tensor types.
	Our implementations of tensor operations are parametrized in terms of multidimensional iterator types and do not rely on specific storage representation of data. The same tensor functions can be utilized for different tensor types including tensor references. Users can extend our framework and provide their own tensor or iterator types with a variety data layouts.
	\item %Tensor template classes support a variety of data layout including the first- and last-order storage format, index offsets, order and dimensions that allows to interface other libraries. Tensor reference classes can be conveniently instantiated with stride-based ranges and used just as tensor types, similar to \tsmall{Matlab}. Both tensor types contain member functions with which tensor applications can be programmed in an object-oriented fashion. 
	We provide tensor template classes with member functions that encapsulate tensor template functions and enable the user to program tensor algorithms in an object-oriented fashion. Numerical results are conveniently verifiable with the \tsmall{\small{Matlab}} toolbox described in~\cite{bader:2006:algorithm862} using overloaded stream operators. Output files can be directly used in the \tsmall{\small{Matlab}} environment without further modification. 
\end{itemize}

%If required, tensor functions of other libraries can be included and conveniently programmed with 
%However, our tensor algorithms do not transform higher-order tensors into matrices such that additional memory space for the unfolded tensors is not needed. In addition, we provide a recursive definition of tensor algorithms due to runtime modifiable parameters such as the mode and extents where a simple nested-loop approach is not sufficient. %with the cost of not being able to use highly optimized basic linear algebra routines. However, we think that the our proposed recursive implementation can be 

%Our framework follows the parametrization principle of the \tsmall{C++} standard library, our tensor algorithms are based on iterator types only. Decoupling algorithms from data structures with the help of iterators allows to have a more orthogonal design.

%\structure{What is the outline of the paper?}
The remainder of the paper is organized as follows. The following Section~\ref{sec:design} provides an overview of our tensor framework and discusses some general design decisions. Section~\ref{sec:tensor} introduces the \tsmall{C++} implementation of our tensor data structures with the focus on the layout and access of elements. We discuss the generality and limitations of our data structure with respect to the storage format. Multidimensional iterators for data structures supporting non-hierarchical data layouts are the topic of Section~\ref{sec:iterator}. Section~\ref{sec:functions} discusses the design and implementation of tensor operations using multidimensional iterators as template parameters. We complete the section by exemplifying the usage of higher-order tensor operations with tensor objects. The last Section~\ref{sec:conclusion} provides a conclusion of this work.

\section{Overview of the Tensor Framework}
\label{sec:design}

%for conveniently implementing numerical multilinear algebra algorithms with dense multidimensional arrays and view. 
%\subsection{Function Scope}
\begin{table}
	\begin{center}
		\begin{tabular}{|lll|} %cp{2.7cm}p{5.5cm}p{2cm}
			\hline
			\tsmall{for\_each}    & \tsmall{transform}    & \tsmall{copy} \\
			\tsmall{fill}         & \tsmall{generate}     & \tsmall{count} \\
			\tsmall{min\_element} & \tsmall{max\_element} & \tsmall{find} \\
			\tsmall{equal}        & \tsmall{mismatch}     & \tsmall{mismatch} \\
			\tsmall{all\_of}      & \tsmall{none\_of}     & \tsmall{any\_of} \\
			\tsmall{iota}         & \tsmall{accumulate}   & \tsmall{inner\_product} \\			
			%\midrule
			\hline
			\tsmall{tensor\_times\_vector}   & \tsmall{tensor\_times\_matrix}   & \tsmall{tensor\_times\_tensor} \\
			\tsmall{tensor\_times\_vectors}  & \tsmall{tensor\_times\_matrices} & \tsmall{transpose}             \\ \tsmall{outer\_product}          & \tsmall{inner\_product} & \tsmall{} \\			
			\hline
		\end{tabular}
	\end{center}
	\caption{\footnotesize Summary of function templates that implement tensor operations for dense tensors and tensor references.}
	\label{tab:summary_tensor_operations}
\end{table}
The primary scope of our framework is given by the following Table~\ref{tab:summary_tensor_operations} that lists basic tensor template functions of the low-level interface of our framework. The first part of the table are first-level tensor operations and correspond semantically to the function templates provided in the algorithm and numeric package of the \tsmall{C++} standard library. Typically, first-level tensor operations process data elements of multidimensional arrays with the same index tuple. The second part are referred to as higher-level tensor operations that are additionally required to implemented numerical multilinear algebra algorithms. Higher-level tensor operations typically manipulate data elements with different index tuples. All functions listed in Table~\ref{tab:summary_tensor_operations} are able to combine multidimensional arrays and views and to process both equally efficiently with same time complexity. %Moreover, our tensor framework allows to utilize the template functions with different implementations of multidimensional arrays and views through iterators.

\subsection{Software Nomenclature}
%\label{sec:design:nomenclature}
In the following we use the nomenclature developed in \cite{stroustrup:2013:cplusplus}. An object is defined in the standard as a region of storage that has a storage duration and a type. The term object type refers to to the type with which an object is created. The \tsmall{C++} language offers fundamental types that are built-in. There are five standard signed and unsigned integer types, a boolean type and three floating point types. We exclude the void type for our discussion. A class is a user-defined type that contains a set of objects of other types and functions that manipulate these objects. A template is a class or a function that we parametrize with a set of types or values. It defines a family of classes or functions or an alias for a family of types. We do not distinguish between the term template class and class template as well as template function and function template. We call the generation of a class or function from a template, template instantiation where the generated template instance is called specialization. Containers are referred to as data structures that manage memory. Algorithms denote template functions that process and manipulate container data with iterators. 

\subsection{Software Design}
The software design of our framework is greatly influenced by the design principle of the Standard Template Library (STL). The STL provides five components algorithms, containers, iterators, function objects and adaptors for this purpose that allows programmers to program data structures of the STL with their own algorithms, and to use algorithms of the STL with their own data structures~\cite{stepanov:1995:standard}. Similarly, the \tsmall{C++} standard library provides multiple containers with different capabilities and runtime complexities, template classes such as the \tsmall{std::vector} that help to organize data~\cite{stroustrup:2013:cplusplus}. Allocation of memory is usually performed with additional predefined or used-defined allocator classes. The \tsmall{C++} standard library also provides free function templates, also known as algorithms, that operate on a one-dimensional range of container data. The range is specified by an iterator pair instantiated by the corresponding container. Iterators of the standard library are one of the five iterator types with different navigation and access capabilities that an algorithm requires for its execution. In other words, an algorithm can only process container data if the algorithm's iterator type requirement is fulfilled by the container's iterator. %As already stated, this approach allows to minimize the number of algorithms in the standard library where one algorithm can process multiple container types with help of iterators. %It allows the library have a high degree of cohesion while providing a loose coupling of software elements. 

We have used this approach to minimize the dependencies between tensor algorithms and the implementation of tensor data structures to yield a separation of concerns. Therefore, we did not use inheritance but mostly composition to establish a loose coupling of software elements. Our tensor framework consists of multiple software stack layers that are illustrated in Figure~\ref{fig:classes}. 
\begin{figure}[th]	
	\centering
	\caption{UML diagram of the software stack of our tensor framework with four components. Black arrows denote associations and visualize the direction of the association. A line annotated with a diamond denotes a composition. Template parameters of the classes are presented with dashed rectangles.}
	\label{fig:classes}
\end{figure}

We will start with the highest layer in the software stack. It consists of two template classes \tsmall{fhg::tensor} and \tsmall{fhg::tensor\_view}. Both provide a high-level interface with which the user can conveniently implement tensor algorithms with member functions. Yet, the implementation of both template classes use software components and call functions of the lower layers. For instance, data organization and access is accomplished with the corresponding template classes \tsmall{fhg::multi\_array} and \tsmall{fhg::multi\_array\_view}, respectively.

The template class \tsmall{fhg::multi\_array} can be regarded as a multidimensional container supporting random access with multiple and single indices. It is designed as a resource handle (similar to the \tsmall{std::vector}) for storing a collection of elements that are ordered in a rectangular pattern. It abstracts from the storage layout in memory and provides a convenient interface to access elements with multi-indices. Allocation and deallocation is handled with the help of the \tsmall{std::vector} template that stores elements contiguously in memory. Its template parameters determine the type of the elements and allocator with which memory is acquired and released. Parameters such as order, dimensions, index offsets or the storage layout are member variables of the \tsmall{fhg::multi\_array} template. Choosing all of the parameters to be runtime-variable allowed us to verify all of tensor template functions and to provide flexible and runtime-adaptable containers.

The \tsmall{fhg::multi\_array\_view} template class is similar to a container adaptor that references a selected region of an \tsmall{fhg::multi\_array} object offering the same functionality as the \tsmall{fhg::multi\_array} template class. It serves as a proxy for conveniently accessing and manipulating selected memory regions of an \tsmall{fhg::multi\_array} object. The template parameter is therefore restricted to a type that has the same properties as \tsmall{fhg::multi\_array}. We have applied the factory method design pattern without subtype polymorphism where an object of type \tsmall{fhg::multi\_array\_view} can only be generated by calling the overloaded function operator of an \tsmall{fhg::multi\_array} object with a tuple of \tsmall{fhg::range} class objects. An \tsmall{fhg::multi\_array} object therefore generates references to itself. 

Both template classes \tsmall{fhg::multi\_array} and \tsmall{fhg::multi\_array\_view} do not provide virtual member functions (especially a virtual destructor) and therefore promote aggregation instead of inheritance. The strategy is also used in the standard library for all containers in order to prevent indirect functions calls and runtime overhead. The template class \tsmall{fhg::tensor} therefore contains the \tsmall{fhg::multi\_array} template class and provies additional member functions with which the user can perform arithmetic operations such as the tensor multiplications. The \tsmall{fhg::tensor\_view} template class is also designed to wrap and extend the functionality of the \tsmall{fhg::multi\_array\_view} template. The element type of the both tensor templates must support scalar arithmetic operations for the unspecialized case. 

The template class \tsmall{fhg::multi\_iterator} defines a multidimensional iterator with which the complete multi-index set of a multidimensional array is accessible while the array can have various data layouts, runtime variable dimensions and order. It provides member functions that create iterator pairs of type \tsmall{iterator\_t} and help to implement tensor functions without explicit specification of the data layout and dimensions. Our framework uses the \tsmall{fhg::stride\_iterator} class for this purpose. The class however does not depend on the \tsmall{fhg::multi\_iterator} template class. It can be instantiated and used separately to modify and access elements of a specific dimension for instance. Both iterator types can be conveniently instantiated with member functions of template classes.

The bottom layer contains tensor template functions that are listed in Table~\ref{tab:summary_tensor_operations}. The template parameters of the function templates denote multidimensional iterator types such as the \tsmall{fhg::multi\_iterator}. The user can utilize our or his own implementation to call the template functions as long as the multidimensional iterator type fulfills specified criteria. Using iterators, the template functions do make assumptions about the underlying data structure or management relieving the user to consider data layout. The tensor template functions are designed in a recursive fashion and implemented as non-member function templates where the maximum depth of the recursion mostly equals the number of dimensions of the corresponding multidimensional array. They do not flatten or rearrange tensors and perform tensor operations in-place using recursion. The framework offers two algorithm packages. The first one contains function templates that have the same semantic and similar function signature as the function templates offered by the algorithm package of the \tsmall{C++} standard library. The second package contains function template that implement tensor multiplication operations such as the tensor-tensor multiplication.

%This is similar to the iterators that have been described in~\cite{garcia:2005:multiarray}. 

%Function templates of both packages are instantiated for the member functions for the \tsmall{fhg::tensor} or \tsmall{fhg::tensor\_view} template classes.

%The implementation is written in \tsmall{C++11} that is defined by the 2011 standard of the \tsmall{C++} language ~\cite{iso:2011:cplusplus}. We have used some of the language features of \tsmall{C++11} that helped us to write our flexible tensor framework with a convenient interface similar to the \tsmall{MATLAB} language. Amongst others, we have used the \tsmall{std::initializer\_list} template class in order conveniently call the constructors of the \tsmall{fhg::range}, \tsmall{fhg::tensor} and \tsmall{fhg::tensor\_view} template classes, the declarator operator \tsmall{\&\&}, i.e. rvalue and universal references and the \tsmall{std::move()} and \tsmall{std::forward()} functions in order to prevent unnecessary copies of temporary objects and variadic templates in order to pass an arbitrary number of arguments for the creation of \tsmall{fhg::tensor\_view} objects.

%\section{Template Classes \texttt{multi\_array} and \texttt{tensor}}
\section{Multidimensional Containers and Views}
\label{sec:tensor}
A multidimensional array is a $p$-dimensional table of values that are arranged in a rectangular pattern and accessible via multiple indices. If it represents higher-order tensors for a given finite basis of the tensor product space, its elements are either real or complex numbers~\cite{lim:2017:hypermatrices, golub:2013:matrix.computations}. For the following discussion, we postulate multidimensional arrays to hold any type of values that can be arranged in a rectangular pattern.
%The following definitions only consider higher-order tensors for a chosen finite basis over the real numbers. %In~\cite{lim:2013:hypermatrices,cichocki:2009:tensor}, higher-order tensors are also described as hypermatrices or multidimensional arrays.
We use the following notation to denote a multidimensional array: 
\begin{equation}
\label{equ:multi.dim.notation}
\mubA = \left (a_{i_1,\dots,i_p} \right )_{i_r \in I_r},
\end{equation}
where $I_r$ is the $r$-th index set with 
\begin{equation}
I_r := \{i_r \in \bbZ \mid o_r \leq i_r < o_r+n_r \ \wedge \ o_r \in \bbZ \wedge n_r \in \bbN \},
\end{equation}
and $|I_r| = n_r$. The number $p$ is a positive integer and will be referred to as the \textit{order} or \textit{rank} of a multidimensional array. The (dimension) extent $n_r$ of the dimension $r$ can be different but must be greater than one. The tuple $\mbn$ of length $p$ is the \textit{shape tuple} of a multidimensional array with $\mbn \in \bbN^p$. Each index $i_r \in I_r$ is biased with an index offset $o_r \in \bbZ$. We denote $\mbo$ with $\mbo \in \bbZ^p$ as the \textit{index offset tuple} of a tensor. We can then derive the \textit{multi-index set} as the Cartesian product of all index sets $I_r$ such that
\begin{equation}
\mcI := I_1 \times I_2 \times \cdots \times I_p,
\end{equation}
with $\mcI \subset \bbZ^p$. We denote an element $(i_1,\dots,i_p)$ of a multi-index set as a \textit{multi-index} with $i_r \in I_r$. Elements of a tensor $\mubA$ are uniquely identifiable using round brackets with a multi-index and
\begin{equation}
\label{equ:multi.dim.notation.element}
\mubA(i_1,i_2,\dots,i_p) = a_{i_1 i_2 \dots i_p}.
\end{equation}
\begin{example}
	Let $\mubA$ be an array of order $p$ with a shape tuple $\mbn$ where $\mbn = (4,2,3)$ and $I_1 = \{0,1,2,3\}$, $I_2 = \{0,1\}$, $I_3 = \{0,1,2\}$. Each element of a tensor can be identified with a multi-index $(i_1,i_2,i_3)$ in $\mcI$. Using the notation in Eq.~\eqref{equ:multi.dim.notation} the multidimensional array can be illustrated as follows:
	\begin{align*}
	\mubA &=
	%\begin{pmatrix}
	\left(\begin{array}{cc|cc|cc}	
	a_{0,0,0} & a_{0,1,0} &  a_{0,0,1} & a_{0,1,1}  & a_{0,0,2} & a_{0,1,2} \\
	a_{1,0,0} & a_{1,1,0} &  a_{1,0,1} & a_{1,1,1}  & a_{1,0,2} & a_{1,1,2} \\
	a_{2,0,0} & a_{2,1,0} &  a_{2,0,1} & a_{2,1,1}  & a_{2,0,2} & a_{2,1,2} \\
	a_{3,0,0} & a_{3,1,0} &  a_{3,0,1} & a_{3,1,1}  & a_{3,0,2} & a_{3,1,2}
	%\end{pmatrix}.
	\end{array}\right)%{cc|cc|cc}
	\end{align*}
\end{example}
A (multidimensional) view  $\mubA'$ of a tensor $\mubA$ is a reference to a specified region or domain of $\mubA$ and has the same order $p$ and data layout $\mbpi$ as the referenced multidimensional array. It can be regarded as a lightweight handle with a shape tuple $\mbn'$ where the dimensions of the view and referenced multidimensional satisfy $n_r' \leq n_r$ for $1 \leq r \leq p$. We define a \textit{section} or \textit{view} of a multidimensional array of order $p$ in terms index-triplets $(f_r, t_r, l_r)$, pairs of indices $(f_r,l_r)$ or scalars $i_r \in I_r$ for all $p$ dimensions. The indices $f_r$, $l_r$ define the lower and upper bound of an index range where $t_r$ the step size for the $r$-th dimension satisfies $0 \geq f_r \geq l_r \geq n_r$ and $t_r \in \bbN$ for $1 \leq r \leq p$. The shape tuple $\mbn'$ of the view $\mubA'$ is given by
\begin{equation}
\label{equ:view_extent}
n_r' = \left \lfloor \frac{l_r - f_r}{t_r} \right \rfloor + 1.
\end{equation}
The $r$-th index set of a view $\mubA'$ is then defined as
\begin{equation}
\label{equ:view_index_set}
I_r' := \{ i_r' \in \bbZ \mid o_r \leq i_r' < n_r'+ o_r \wedge o_r \in \bbZ\},
\end{equation}
where $o_r$ is the $r$-th index offset of the referenced multidimensional array $\mubA$. Using the notation in the previous section, a view $\mubA'$ of a multidimensional array $\mubA$ is denoted by
\begin{equation}
\mubA' = \left( a_{i_1',\dots,i_p'}'\right)_{i_r' \in I_r'}.
\end{equation}
Analogous to the multidimensional arrays, the multi-index set of a view is given by
\begin{equation}
\label{equ:multi_index_set_view}
\mcI' := I_1' \times I_2' \times \cdots \times I_p',
\end{equation}
where $\mcI' \subseteq \mcI \subseteq \bbZ^p$. A \textit{slice} is a special kind of view with two dimensions that have the same extent as the referenced multidimensional array. Let $k$ and $l$ be the subscripts of those two dimensions. The resulting multi-index set $\mcI'$ of the slice is given by Eq.~\eqref{equ:multi_index_set_view} with
\begin{equation}
I_r = 
\begin{cases}
\{o_r,\dots,o_r+n_r\} &\ \text{for}\ r = k \wedge r = l, \\
\{o_r\} &\ \text{otherwise.}
\end{cases}
\end{equation}
A \textit{fiber} has exactly one dimension that is greater than one. 
\begin{example}
	Let $\mubA$ be the $3$-way array from the previous example and zero-based index offsets. Let $f_1 = 1$, $l_1 = 3$, $t_1 = 2$, and $f_2 = 0$, $l_2 = 1$ and $f_3 = l_3 = 2$ index triplets for choosing a section of $\mubA$. With $\mbn' = (2,2,1)$ being the shape tuple of the view $\mubA'$, elements of the view are given by 
	\begin{equation*}
	\mubA' = 
	\begin{pmatrix}
	a_{0,0,0}' & a_{0,1,0}' \\
	a_{1,0,0}' & a_{1,1,0}'
	\end{pmatrix} = 
	\begin{pmatrix}
	a_{1,0,2} & a_{1,1,2} \\
	a_{3,0,2} & a_{3,1,2}
	\end{pmatrix}.
	\end{equation*}
\end{example}
%Tensors will be denoted with underscored bold capital letters such as $\mubA$ or $\mubC$. Two-dimensional and one-dimensional tensors will be called matrices and vectors, respectively. Matrices are represented without an underscore $\mbU$ or $\mbV$ and vectors or tuples are given by small bold letters such as $\mbu$. We will call a small bold letter as a tuple if its elements are either natural numbers or integers. Letters that are not bold and small are scalars. 

%Please consider the following basic operations for tensors that have been also described in~\cite{lim:2013:hypermatrices,cichocki:2009:tensor}.

\subsection{Data Organization and Layout}
The above notation of a tensor is an abstract representation of a multidimensional array. Usually a multidimensional array stores its elements in memory where the latter can be addressed with a single index. We refer to that single index $k$ as the (absolute) memory address. Our implementation of the tensor template class is stored contiguously in memory such that elements of array are addressable using a single index $j$ as well. The single index shall denote a relative position within the multidimensional array. In that case, the $j$-th element of $\mubA$ in memory is then given by the linear function
\begin{equation}
\label{equ:absolute.memory.address}
k = k_0 + j \cdot \delta,
\end{equation}
where $\delta$ is the number of bytes to store an element of a multidimensional array $\mubA$ and $k_0 \in \bbN_0$ is the (absolute) memory address of the first element. The domain of the function is given by
\begin{equation}
\label{equ:plain.index.space}
\mcJ := \{0,1,\dots, \prod_{r=1}^p n_r-1\}
\end{equation}
which we denote as the \textit{(relative) memory index set} of a multidimensional array of order $p$ where $n_r$ is the length of the $r$-th dimension. Therefore, we may enumerate elements of a multidimensional array $\mubA$ with a single scalar index $j$ of the index set $\mcJ$. We denote elements of the memory index set as \textit{memory indices} of the multidimensional array that correspond to displacements relative to the memory address of the first array element.

Given a contiguous region within a single indexed memory space, the \textit{data layout} (storage format) of a contiguously stored dense multidimensional array defines the ordering of its elements within the specified memory region. A multidimensional array with the dimensions $n_1,\dots,n_p$ has $\prod_r n_r!$ possible orderings. In practice however, only a subset of all possible orderings are considered. In case of two dimensions for instance, most programming languages arrange elements of two-dimensional arrays either according to the row- or column-major storage format where adjacent elements of a row or column are successively stored in memory. More sophisticated non-linear layout functions have been investigated for instance in~\cite{chatterjee:1999:recursive, elmroth:2004:recursive} with the purpose to increase the data locality of dense matrix operations. These type of layout functions partition two-dimensional arrays into hierarchical ordered blocks where elements of a block are stored contiguously in memory. 

Our data structure supports all non-hierarchical data layouts including the first- and last-order storage formats. We denote both formats as standard data layouts of dense multidimensional arrays, where the former format is defined in the \tsmall{Fortran}, the latter in the \tsmall{C} and \tsmall{C++} language specification, respectively. Non-hierarchical data layouts can be expressed in terms of permutation tuples $\mbpi$ which we denote as the \textit{(data) layout tuple}. The $q$-th element $\pi_q$ of a permutation tuple $\mbpi$ corresponds to an index subscript $r$ of a multi-index $i_r$ with the precedence $q$ where $i_r \in I_r$ and $1 \leq q,r \leq p$. % Please note that $\mbpi(q)$ corresponds to the same index subscript as $\pi_q$.
In case of the first-order format, the layout tuple is defined as
\begin{equation}
\label{equ:layout.tuple.first.order}
\mbpi_F := (1,2,\dots,p),
\end{equation}
where the precedence of the dimension ascends with increasing index subscript. The layout tuple of the last-order storage format is given by 
\begin{equation}
\label{equ:layout.tuple.last.order}
\mbpi_{L} := (p,p-1,\dots,1).
\end{equation}
We might therefore interpret the layout tuple also as a precedence tuple of the dimensions. The set of all possible layout tuples is denoted by $\Pi_p$. The number of all possible non-hierarchical element arrangements is $p!$ and equal to the number of elements in $\Pi_p$.

The $q$-th stride $w_q$ is a positive integer and defined as the distance between two elements with identical multi-indices except the $q$-th indices differing by one. More specifally, given a layout tuple $\mbpi$ and the shape tuple $\mbn$, elements of a stride tuple $\mbw$ are given by:
\begin{equation}
\label{equ:stride.tuple.permutation}
w_{\pi_r} = 
\begin{cases}
1 &\quad \text{for} \ r = 1.\\
\prod_{q=1}^{r-1} n_{\pi_q} &\quad \text{otherwise}.
\end{cases}
\end{equation}
Please note that for any layout tuple $\mbpi$ the corresponding stride tuple $\mbw$ satisfies $1 \leq w_{\pi_q} \leq w_{\pi_r}$ with $1\leq q < r \leq p$. For instance, the \tsmall{Fortran} language specification stores elements of a multidimensional array according to the first-order storage format such that the corresponding stride tuple is defined as
\begin{equation}
\label{equ:stride.tuple.first.order}
\mbw_{F} = (1,\ n_1,\ n_1\cdot n_2,\ \dots ,\ \prod_{r=1}^{p-1} n_r).
\end{equation}
In case of the first-order storage format, the $q$-th stride increases with consecutive index subscripts by the factor of $n_{q-1}$. The last-order storage format is given by $\mbpi_{L} = (p,p-1,\dots,1)$ Given the layout tuple in Eq.~\eqref{equ:layout.tuple.last.order}, the stride tuple is then 
\begin{equation}
\label{equ:stride.tuple.last.order}
\mbw_{L} = (\prod_{r=2}^p n_r, \prod_{r=3}^p n_r, \dots, n_p,1),
\end{equation}
which corresponds to the definition in the 2011 C-language specification. We denote the set of all stride tuples by $\mcW$ with $\mcW \subseteq \bbN^p$. 

%\subsection{Auxiliary Class} 
%The template classes \tsmall{fhg::shape}, \tsmall{fhg::layout} and \tsmall{fhg::offset} are auxiliary classes with which the \tsmall{fhg::tensor} is instantiated. 
%The instantiation of the \tsmall{fhg::tensor\_view} requires objects of type \tsmall{fhg::span} that define a range for an extent of the selected \tsmall{fhg::tensor} object.
%\todo{Erzählen wie diese instantiiert werden und wie sie der Klasse helfen.}

\subsection{Tensor Template Class} 
The parameterized class \tsmall{fhg::tensor} is a sequence container which is responsible for the data management. It resembles the vector template class \tsmall{std::vector} of the \tsmall{C++} standard library. It has two template parameters, \tsmall{value\_type} and \tsmall{allocator}. The former one determines the data type of the elements, the latter the type of the memory allocator. The type \tsmall{value\_type} must be a numeric type with which addition, subtraction, multiplication and division can be performed. 
\begin{lstlisting}
template <class value_type, class allocator>
class tensor;
\end{lstlisting}
This restriction arises due to member functions such as overloaded arithmetic operators implementing pointwise tensor operations for convenience. In order to support a wider range of applications we offer the template class \tsmall{fhg::multi\_array} that does not include any arithmetic operations and therefore has lesser restrictions on the data type of its elements. It has the same template parameters and member variables as the \tsmall{fhg::tensor} template class. The user is for instance free to set the first template parameter to boolean data type and equip the \tsmall{fhg::multi\_array} with bitwise operations. 

Similar to the \tsmall{std::vector} template class, \tsmall{fhg::tensor} contains public member types such as \tsmall{value\_type}, \tsmall{size\_type}, \tsmall{difference\_type}, \tsmall{pointer}, \tsmall{const}\tsmall{\_pointer}, \texttt{reference}, \tsmall{const}\allowbreak\tsmall{\_reference}, \tsmall{iterator}, \tsmall{const}\allowbreak\tsmall{\_iterator}. The last two types denote multidimensional iterators which will be explained in the following subsection. The template class also defines the member type \tsmall{tensor\_view\_t} that is \tsmall{fhg::tensor\_view<tensor>} where tensor is the data structure itself. 
%Additionally, \tsmall{fhg::range} is a template class instance of \tsmall{fhg::basic\_range} where the template parameter is \tsmall{std::ptrdiff\_t}. 
Consider the following listing with template class instances of the template class \tsmall{fhg::tensor}.
\begin{lstlisting}[classoffset=1,morekeywords={cftensor,dtensor,itensor}]
using cftensor = fhg::tensor<std:complex<float>>;
using itensor  = fhg::tensor<int>;
using dtensor  = fhg::tensor<double>;
\end{lstlisting}
Instances of the first specialization contain complex numbers in single precision, objects of the second type integer numbers and the third numbers in double precision. The template class \tsmall{tensor} provides public member functions in order to 
\begin{itemize}
	\item instantiate and copy tensor instances, e.g. \tsmall{tensor()},
	\item assign data elements, e.g. \tsmall{operator=(..)},
	\item access elements, e.g. \tsmall{at(..)}, \tsmall{operator[..]},
	\item perform tensor multiplication operations, e.g. \tsmall{times\_tensor(..)}, \tsmall{inner(..)},
	\item perform tensor pointwise operations, e.g. \tsmall{operator+=(..)}, \tsmall{operator-()},
	\item generate views, e.g. \tsmall{operator()(..)},
	\item generate iterators, e.g. \tsmall{begin()}, \tsmall{mbegin()},	
\end{itemize}
including all size and capacity functions of the \tsmall{std::vector} such as \tsmall{size()}, \tsmall{empty()}, \tsmall{clear()}, \tsmall{data()}. However, we do not support stack and list operations such as \tsmall{push\_back}, \tsmall{pop\_back}, \tsmall{insert} or \tsmall{erase}. 

%\subsubsection{Constructors and Assignments}
The type \tsmall{fhg::tensor} shall denote the type \tsmall{fhg::tensor}\allowbreak\tsmall{<value\_type,}\tsmall{allocator>} for given template parameters \tsmall{value\_type} and \tsmall{allocator}. The \tsmall{fhg::tensor} template class provides the following constructor declarations.
\begin{lstlisting}[classoffset=1,morekeywords={Extents}]
tensor();
tensor(shape const&);
tensor(shape const&, offset const&);
tensor(shape const&, layout const&);
tensor(shape const&, offset const&, layout const&);
\end{lstlisting}
The default constructor of the \tsmall{fhg::tensor} class calls the default constructor of all member variables. The \tsmall{fhg::tensor} template class contains the following \tsmall{std::vector} private member variables in order to store
\begin{itemize}
	\item data elements of type \tsmall{value\_type},
	\item extents of type \tsmall{size\_type},
	\item layout elements of type \tsmall{size\_type},
	\item strides of type \tsmall{size\_type},
	\item index offsets of type \tsmall{difference\_type}.
\end{itemize}
The above mentioned public member types are derived from the \tsmall{std::vector} template class with the specified template parameters \tsmall{value\_type} and \tsmall{allocator}. The member variable of type \tsmall{std::vector}\allowbreak\tsmall{<value\_type}\allowbreak\tsmall{,allocator>} helps \tsmall{fhg::tensor} class template to allocate and free memory space where the data elements are contiguously stored. The position of the elements corresponds to indices of the plain index set given by Eq.~\eqref{equ:plain.index.space}. The second constructor specifies the extents with the help of \tsmall{fhg::shape} class instances. The order is automatically derived and the length of the data vector is determined. The default layout tuple is computed with Eq.~\eqref{equ:layout.tuple.first.order}. %For non-hierarchical data layouts, a \textit{stride tuple} $\mbw$ with strides $w_r$ greater or equal one suffices to define the spacing between all elements of the multidimensional array in correspondence with the data layout. 
The constructor can then compute the strides with Eq.~\eqref{equ:stride.tuple.permutation} initializing the index offsets with zero. The remaining constructors allow to specify different offset and/or layout tuples using the auxiliary \tsmall{fhg::layout} and \tsmall{fhg::offset} classes, respectively. The specification of the layout using the \tsmall{fhg::layout} class is constructed with the a permutation tuple with one-based indices. All three auxiliary classes provide an initialization with the \tsmall{std::initializer\_list} for a convenient initialization of their elements. The elements are stored with a specialization of the \tsmall{std::vector} template class. Consider the following example where two statements instantiate objects of the \tsmall{fhg::tensor<double>} class.
\begin{example}
\label{ex:tensor}
The first constructor generates an object \tsmall{A} where the offsets are all zero and the elements are stored according to the first-order storage format and therefore initializing the layout tuple with \tsmall{\{1,2,3\}} as given by Eq.~\eqref{equ:layout.tuple.first.order}. The second object \tsmall{B} has the same shape but initializes the index offsets with \tsmall{1,-1, 0} using the last-order format for the element ordering.
\begin{lstlisting}
tensor<double> A(shape {4,2,3});
tensor<double> B(shape {4,2,3}, offset {1,-1,0}, layout {3,2,1});
\end{lstlisting}
We can therefore assert that the strides of objects \tsmall{A} and \tsmall{B} are \tsmall{\{1,4,8\}} and \tsmall{\{6,2,1\}}, respectively.
\end{example}

The copy assignment operators \tsmall{operator=()} of the \tsmall{fhg::tensor} class template are responsible for copyin data and protecting against self-assignment. Consider the following overloaded assignment operators.
\begin{lstlisting}
tensor& operator=(tensor        const&);
tensor& operator=(tensor_view_t const&);
\end{lstlisting}
%tensor& operator=(std::vector<value_type> const&);
%tensor& operator=(const_reference);
The first method copies all private member variables of the source instance such that source and destination have the same shape tuple and the same order after the assignment. The layout, stride and index offset tuples of the destination tensor object are only copied if the order of the source and destination tensor objects differ. In either cases, the user can expect the source and destination \tsmall{fhg::tensor} class instances to be equal and independent after the copy operation, see~\cite{stroustrup:2013:cplusplus} for a detailed discussion. 
%TODO: What happens if A = Aview or A = A? , ~\cite[p.510]{stroustrup:2013:cplusplus}
We defined two tensor to be equal if they have the shape tuple, order and elements with the same multi-index independent of their layout, index offset tuple and allocator. The second assignment operator copies all elements of the source \tsmall{fhg::tensor\_view} class template instance including the shape tuple. The layout and index tuples are only copied if the order to the source and destination tensor objects differ. 
\begin{example}
\label{ex:relayout}
Let \tsmall{A} and \tsmall{B} be \tsmall{fhg::tensor} objects of the previous Example~\ref{ex:tensor}. The following code snippet initializes all elements of \tsmall{A} with the value one using the assignment operator and copies the content of \tsmall{A} to \tsmall{B}. 
\begin{lstlisting}[classoffset=1,morekeywords={assert}]
B = A = 1;
\end{lstlisting}
As both objects do have equal extents, only data values of \tsmall{A} are copied to \tsmall{B} without modifying the offset and layout tuple of \tsmall{B}.
\end{example}
Besides the type of the data elements, the user can change the content and the size of all member variables at runtime. The framework offers \tsmall{reshape} and \tsmall{relayout} member functions that allow to dynamically adjust the shape and layout tuple, respectively. The container automatically adjusts the strides and reorders the in memory if necessary.
The \tsmall{fhg::tensor} template class provides member functions for accessing elements with multi-indices and scalar memory indices. The following listing depicts two member functions that return a \tsmall{reference} of the specified element.
\begin{lstlisting}[classoffset=1,morekeywords={MultiIndex}]
template<class ... MultiIndex>
reference at(MultiIndex&& ...);
reference operator[](size_type index);
\end{lstlisting}
The first access function is a template function with a variadic template that allows to conveniently access elements with multi-indicies. It transforms multi-indices onto memory indices according to the stride tuple of the template class. Given the order $p$ and a stride tuple $\mbw$ with respect to a layout tuple $\mbpi$ and stride tuple $\mbw$ given by Eq.~\eqref{equ:stride.tuple.permutation}, the \textit{non-hierarchical layout function} $\lambda$ is defined by
\begin{equation}\label{equ:f}
\lambda : \bbN^p \times \bbZ^p \times \bbZ^p \rightarrow \bbN, \quad \lambda(\mbw,\mbi,\mbo) = \sum_{r=1}^p w_r (i_r - o_r).
\end{equation}
For fixed stride tuples $\mbw_F$ and $\mbw_L$, the layout functions $\lambda_{\mbw_F}$ and $\lambda_{\mbw_L}$ coincide with the definitions provided in~\cite{chatterjee:1999:recursive, andres:2010:runtime, garcia:2005:multiarray}. %Furthermore, for two-dimensional arrays, the layout function $\lambda_{\mbw_F}$ corresponds to the row-major and $\lambda_{\mbw_L}$ to the column-major layout. % In that case index pairs $(i,j)$ specifying elements $a_{ij}$ of a two-dimensional array with $m$ rows and $n$ columns, are mapped either by the row-major $\lambda_r$ or column-major layout function $\lambda_c$ onto relative memory indices according to the equation $\lambda_r(i,j) = i + m \cdot j$ and $\lambda_c(i,j) = j + n \cdot i$. 
\begin{example}
\label{ex:superdiagonal_array}
Let \tsmall{A} be a multidimensional array of order 3 and type \tsmall{fhg::tensor} with equal extents $n$ and elements all set to zero. We can create an identity tensor with ones in the superdiagonals by the following statement.
\begin{lstlisting}
for(auto i = 0u; i < A.extents().at(0); ++i)
	A.at(i,i,i) = 1.0;
\end{lstlisting}
Object \tsmall{A} is processed independently of its layout tuple. 
\end{example}
The argument size of the variadic template however must be specified at compile time. Note that, the multidimensional array \tsmall{A} of Example~\ref{ex:superdiagonal_array} must be of order three. The class offers additional member functions that allow to specifiy multi-indices with the runtime-variable length using the \tsmall{std::vector}. Consider the following code snippet in the next example.
\begin{example}
Given a multidimensional array \tsmall{A} of order $p$ where $p$ is initialized at runtime. We can create an identity tensor with a vector of size $p$ that is initialized with the value $i$.
\label{ex:superdiagonal_array_2}
\begin{lstlisting}
for(auto i = 0u; i < A.extents().at(0); ++i){
	A.at(std::vector<std::size_t>{p,i}) = 1.0;
}
\end{lstlisting}
\end{example}
Using multi-indices abstracts from the underlying data layout and enables the user to write layout invariant programs as all elements have a unique multi-index independent of the data layout, see Eq.~\eqref{equ:multi.dim.notation.element}. However, each element access with multi-indices involves a multi-index to memory index transformation that is given by Eq.~\eqref{equ:f}.

The second member function of the previous listing provides memory access with a single scalar parameter corresponding to memory indices of the memory index set of a multidimensional array, see Eq.~\eqref{equ:plain.index.space}. Consider the following example. 
\begin{example}
Let \tsmall{A} be a multidimensional array specified in the previous example. The statement in the listing sets all elements of the object using a single induction variable \tsmall{j} that takes values of the complete memory index set of the array. 
\begin{lstlisting}
for(auto j = 0u; i < A.size(); ++j)
	A[j] = 0.0;
\end{lstlisting}
\end{example}
In contrast to the access with multi-indices, referencing elements with memory indices does not include index transformations. It might be seen as a low-level access where the user is responsible for referencing the correct elements. We might say that scalar indexing is convenient whenever the complete memory index set is accessed and the multi-index or the ordering of the data elements are not relevant for the implementation of the tensor operation. Higher-level tensor operations such as the tensor transposition require some type of multi-index access.

%%%%%%%%%%%%%%%%%%%%%%%%%%%%%%%%%%%%%
%%%%%%%%%%%%%%%%%%%%%%%%%%%%%%%%%%%%%
%%%%%%%%%%%%%%%%%%%%%%%%%%%%%%%%%%%%%

\subsection{View Template Class}
\label{sec:tensor:view}
An instance of the parameterized class \tsmall{fhg::tensor\_view} references a container instance that has been selected using \tsmall{fhg::range} class instances. 
\begin{lstlisting}[classoffset=1,morekeywords={tensor_t}]
template <class tensor_t>
class tensor_view;
\end{lstlisting}
The template parameter \tsmall{tensor\_t} can be of type \tsmall{fhg::tensor<value\_type, allocator>}. However, the user can also set a different \tsmall{tensor\_t} type. In that case \tsmall{tensor\_t} type requires the type \tsmall{tensor\_t::value\_type} to be an arithmetic type and the referenced container to store its data contiguously with access to all previously mentioned public member types and variables. A template class instance \tsmall{fhg::tensor\_view} contains therefore the same public member types and member methods as the referenced template class instance \tsmall{fhg::tensor} with only minor differences. This has the advantage that both template class instances \tsmall{fhg::tensor} and \tsmall{fhg::tensor\_view} can be treated almost identically.

The \tsmall{fhg::tensor\_view} template class contains the following private member variables:
\begin{itemize}
	\item a pointer to the selected tensor of type \tsmall{tensor\_t*},
	\item a pointer to the first element of type \tsmall{value\_type*},
	\item ranges of type \tsmall{fhg::range},
	\item extents of type \tsmall{size\_type},
	\item strides of type \tsmall{size\_type}.
\end{itemize}
The layout and offset tuple can be obtained by calling the appropriate functions of the referenced \tsmall{fhg::tensor} object. The constructor only requires the specification of the referenced \tsmall{fhg::tensor} and $p$ \tsmall{fhg::range} objects that are used to initialize the remaining private member are initialized. The $r$-th \tsmall{fhg::range} object defines an index set $I_r'$ that must be a subset of the index set $I_r$ of the \tsmall{fhg::tensor} as we have already previously defined, see Eq.~\eqref{equ:view_index_set}. Moreover, the constructor of the \tsmall{fhg::range} class takes either two or three indices where \tsmall{first}, \tsmall{last} and \tsmall{step} corresponds to $f_r$, $l_r$ and $t_r$ indices of Eq.~\eqref{equ:view_index_set}.
\begin{lstlisting}[classoffset=1,morekeywords={size_type}]
range(size_type first, size_type last);
range(size_type first, size_type step, size_type last);
\end{lstlisting}
The constructor also tests if the specified ranges are valid and do not violate any bounds of the selected \tsmall{tensor\_t} instance. The extents $n_r'$ of the view are therefore computed according to Eq.~\eqref{equ:view_extent}. The pointer to the first element of the array is given by adding an offset $\gamma$ to the pointer to first element of the selected \tsmall{fhg::tensor}. The offset is computed by combining the $p$ lower bounds \tsmall{first} using the layout function $\lambda$ in Eq.~\eqref{equ:f} such that
\begin{equation}
\label{equ:view_gamma}
\gamma := \lambda(\mbw,\mbf,\mbo), \quad \text{with}\ \mbf = (f_1,\dots,f_p) 
\end{equation}
where $\mbw$ and $\mbo$ are the stride and offset tuple of the \tsmall{fhg::tensor} object and $f_r$ is the lower bound \tsmall{first} of the $r$-th \tsmall{fhg::range} instance. The stride tuple $\mbw'$ of a view can be computed with Eq.~\eqref{equ:stride.tuple.permutation}.

The tensor template class \tsmall{fhg::tensor} provides two methods that overload the function call operator and generate class instances of type \tsmall{fhg::tensor\_view<tensor>}. The latter type is equivalent to \tsmall{tensor\_view\_t} where \tsmall{tensor} is the type of the calling template class instance. Please consider the following listing which is an excerpt of the \tsmall{fhg::tensor} template class.
\begin{lstlisting}[classoffset=1,morekeywords={domain,range}]
template<class ... domain>
tensor_view_t operator()(domain&& ...);
tensor_view_t operator()(std::vector<range> const&);
\end{lstlisting}
The first function is a template function with a template parameter pack \tsmall{domain} that is recursively unpacked for either integers or objects of type \tsmall{fhg::range}, i.e. index triplet or pairs. The unpacking however is performed with protected template functions that are called from the constructor of the corresponding \tsmall{tensor\_view\_t} class. 
\begin{lstlisting}[classoffset=1,morekeywords={tensor_t}]
tensor_view(tensor_t*);
template<class ... domain>
tensor_view(tensor_t*, domain&& ...);
\end{lstlisting}
Both methods are declared as friend in the \tsmall{tensor\_view} template class and only callable from the referenced \tsmall{fhg::tensor} objects. The first constructor generates a reference to an unselected \tsmall{tensor\_t} object where no domain is specified. This means that the reference covers the complete domain of the tensor object that has been selected. The second constructor uses a variadic template that processes either integers or objects of type \tsmall{fhg::range}. Both constructors are called from the tensor template class from the overloaded function call operator. Please note that \tsmall{fhg::tensor\_view} the class instance cannot be instantiated without a \tsmall{tensor\_t} object as the default constructor of \tsmall{fhg::tensor\_view} the template class is deleted. Consider the following example that illustrates the creation of an view object \tsmall{Av} of type \tsmall{tensor\_view<tensor<float>{}>}. Please note that only a \tsmall{tensor\_t} object is able to call the protected \tsmall{fhg::tensor\_view} constructor with a function call operator of \tsmall{tensor\_t}.
\begin{example}
\label{ex:view}
Let \tsmall{A} be an object tensor of type \tsmall{tensor<float>} from the previous Example~\ref{ex:tensor} with \tsmall{n = \{4,2,3\}} and zero-based index offsets. Let also \tsmall{r1 = range(1,2,3)}, \tsmall{r2 = range(0,1)} and \tsmall{r3 = range(2)} be index ranges with which the $3$-way array \tsmall{A} is selected. According to Eq.~\eqref{equ:view_extent} the shape tuple of the view \tsmall{Av} is then given by \tsmall{nv = \{2,2,1\}}.
\begin{lstlisting}
tensor_view<tensor<float>> Av(A.data(), range{1,2,3}, range{0,1}, 2);
\end{lstlisting}
The $3$-way array \tsmall{A} can also be seen as a composition of three slices \tsmall{Av0}, \tsmall{Av1} and \tsmall{Av2} where the \tsmall{i}-th slice is given by
\begin{lstlisting}
tensor_view<tensor<float>> Avi(A.data(), range{}, range{}, i);
\end{lstlisting}
where an empty range object contains all indices of the corresponding index set.
%whereas $\mba_{i_2,i_3}$ corresponds to the $1$-mode $(i_2,i_3)$-th fiber with shape tuple $(n_1)$ of the multidimensional array. %\todo[noline]{Zitiere hier aufgrund der Namensgebung Fiber and Slice.}
\end{example}
The instantiation of the view \tsmall{Av} is only exemplified in the following listing in order to demonstrate the instantiation procedure. Please note that we did not provide a mechanism for counting the number of \tsmall{fhg::tensor\_view} objects associated to one \tsmall{tensor\_t} object. The implementation of such a mechanism requires tensor objects to know about their references and to signal its views when the destructor is called. An existing \tsmall{fhg::tensor\_view} object can therefore become invalid when its referenced \tsmall{fhg::tensor} instance does not exist any more. This situation is similar to a dangling pointer. The user is therefore responsible for avoiding the situation where the referenced \tsmall{fhg::tensor} object falls out of scope or is deleted. An alternative way to implement views of tensors is to only allow temporary \tsmall{fhg::tensor\_view} instances that is an  \tsmall{rvalue}. In order to enable this restriction, the copy constructor and copy assignment operator needs to be deleted or hidden from the user. In that case, only functions that allow only \tsmall{rvalue} references of \tsmall{fhg::tensor\_view} instances using the \tsmall{\&\&} declarator operator. Please consider following listing that demonstrates the instantiation of tensor views.
\begin{example}
\label{ex:createview}
Let \tsmall{A} be an instance of \tsmall{fhg::tensor<float>} template class with the shape tuple \tsmall{(3,4,2)}. We can then create a view \tsmall{Av} of \tsmall{A} by calling the overloaded call operator \tsmall{operator()} of \tsmall{A} with \tsmall{fhg::range} instances.
\begin{lstlisting}
tensor_view<tensor<float>> Av = A ( range {1,2}, range {1,2}, 1 );
\end{lstlisting}
\end{example}
The operator calls the constructor of the \tsmall{tensor\_view\_t} class with the specified ranges. Please note that instead of using the \tsmall{auto} specifier, we explicitly defined the type of the view for demonstration purposes. The number of arguments in the parameter pack of the template function must be known at compile time. If the number of \tsmall{fhg::range} instances are runtime-variable, the user can call the second member function that creates a view of tensor using a \tsmall{std::vector}.  Please note that statement in Example~\ref{ex:createview} is very similar to the \tsmall{MATLAB} syntax where a section of a multidimensional array \tsmall{A} is created with \tsmall{A(1:2,1:2,1)} using the call operator in conjunction with the colon operator.

Access functions of the \tsmall{fhg::tensor\_view} template class with multi-indices depict the same interface as for the \tsmall{fhg::tensor} template class.
\begin{lstlisting}[classoffset=1,morekeywords={MultiIndex}]
template<class ... MultiIndex>
reference at(MultiIndex&& ...);
reference operator[](size_type);
\end{lstlisting}
However, the access function cannot use the layout function $\lambda$ in Eq.~\eqref{equ:f} in order to generate memory indices. Each index in $I_r'$ needs to be transformed into an index of the set $I_r$ before an element can be accessed. Memory indices of a view's elements are then given by the function $\lambda'$ with
\begin{equation}
\label{equ:lambda_view}
\lambda' :  \bbN^p \times \bbZ^p \times \bbZ^p \rightarrow \bbN, \quad \lambda'(\mbw',\mbi',\mbo) = \gamma + \lambda(\mbw'',\mbi',\mbo)
\end{equation}
where $\gamma$ is the offset given by Eq.~\eqref{equ:view_gamma} and $\mbw''$ is a modified stride tuple with $w_r'' = w_r' t_r$.
\begin{example}
\label{ex:superdiagonal_view}
Let \tsmall{A} be the multidimensional array where all the extent of the dimensions are \tsmall{4}. Similar to the Example~\ref{ex:superdiagonal_array}, we can instantiate all diagonal elements of a slice \tsmall{Av} to one using a single for-loop.
\begin{lstlisting}
auto Av = A ( range {}, range {}, 1 );
for(auto i = 0; i < 4; ++i)
	Av.at(i,i,0) = 1;
\end{lstlisting}
Internally, the \tsmall{at()} operation computes the relative memory indices from the multi-indices \tsmall{(i,i,0)} according to Eq.~\eqref{equ:lambda_view}.
\end{example}
The instantiation of multidimensional iterators, views and the interface will be postponed to the following chapters.

\subsection{Output Stream Class}
In order to compare and verify our numerical results with \tsmall{MATLAB} we provided our own output stream class \tsmall{fhg::matlab\_ostream} for a formatted output of the \tsmall{fhg::tensor} and \tsmall{fhg::tensor\_view} instances. The class overloads stream operators \tsmall{{<}{<}} for the formatted output. The constructor of the class  expects an instance of the \tsmall{std::basic\_ios} class template from the standard template library such as \tsmall{std::cout}. The user can directly utilize the instance \tsmall{fhg::mcout} of the \tsmall{fhg::matlab\_ostream} for convenience. The stream function relayouts the multidimensional array for the first-order storage format and outputs the elements of the tensor using the \tsmall{MATLAB} notation. One can also input additional \tsmall{MATLAB} commands into the stream such as plotting statements. In this way, \tsmall{MATLAB} scripts can be generated with our tensor library. Consider the following listing where the complete tensor \tsmall{A} is inserted into the standard output stream. 
\begin{example}
The elements of \tsmall{A} are stored according to the last-order storage format and their values are initialized with indices of the plain index set.
\begin{lstlisting}[classoffset=1,morekeywords={ostream,matlab_ostream,mcout}]
fhg::mcout << "A = " << A << std::endl;
fhg::mcout << "plot(A(:) - Aref(:));" << std::endl;
// output:
// A = cat(3, [ 0 2 4 6 ; 8 10 12 14 ; 16 18 20 22 ],...
//            [ 1 3 5 7 ; 9 11 13 15 ; 17 19 21 23 ]);
// plot(A(:) - Aref(:));
\end{lstlisting} %
\end{example}
The output can be directly copied into \tsmall{MATLAB}'s command window and executed if the reference tensor \tsmall{Aref} is already defined. Please note that the \tsmall{matlab\_stream} object could have been also instantiated with an output file stream \tsmall{std::ofstream}. We have used this approach to verify the results of our tensor algorithms with the tensor toolbox described in~\cite{bader:2006:algorithm862}. The framework also provides overloaded input and output stream operators for the tensor class template in conjunction with standard output streams.

%The source file \textit{example2.cpp} contains instantiations and initializations of the \texttt{tensor} template class using different layout tuples. 
%\todo[noline]{More on input and output streams.}
\begin{comment}
\begin{lstlisting}
std::cout << fhg::fmt:simple << A << std::endl;
// 0 1 2 3 4 5 6 7 8 9 ... 23
\end{lstlisting}
\begin{lstlisting}
std::cout << fhg::fmt:matlab << A << std::endl;
// A = cat(3, [ 0 2 4 6 ; 8 10 12 14 ; 16 18 20 22 ],...
//            [ 1 3 5 7 ; 9 11 13 15 ; 17 19 21 23 ]);
\end{lstlisting}
\begin{lstlisting}
std::cout << layout::matlab << A << std::endl;
// 0 1 2 3 4 5 6 7 8 9 ... 23
\end{lstlisting}
\end{comment}

\section{Multidimensional Iterator}
\label{sec:iterator}
An iterator is data structure that represents positions of elements in a container and allows to traverse between successive elements of the container. The iterator concept allows to decouple or minimize the dependency between algorithms that use iterators and data structures that create iterators. Pointers for instance are akin to iterators and may be regarded as an instance of the iterator concept. A pair of iterators $[a,b)$ define a (half-open) range of the referenced container. Containers of the standard template library provide member functions \tsmall{begin()} and \tsmall{end()} in order to generate a half-open range for the corresponding container where the iterator $a$ points to the first element and $b$ to the position after the last element.

The standard template library divides an iterator into five iterator categories with different capabilities where random-access iterators provide the largest number of access and iteration methods with iterator arithmetic similar to that of a pointer. Moreover, it accesses any valid memory locations in constant time. Template functions provided by the standard template library operate with iterators and determine the required category of the iterators. The \tsmall{std::for\_each()} template function for instance works with input iterators while \tsmall{std::sort()} requires iterators with a random-access iterator tag. Hence, not every container can instantiate a random-access iterator such as associative container \tsmall{std::map} such that \tsmall{std::sort()} is not compatible with iterators generated by \tsmall{std::map} objects. However, iterator instances generated by \tsmall{std::vector}, \tsmall{std::array} and \tsmall{std::deque} objects support random-access and can use the function \tsmall{std::sort()}. In case of the containers \tsmall{std::vector}, \tsmall{std::array} and \tsmall{std::deque} that implement one-dimensional array with a contiguous memory region, the half-open range covers the complete (memory) index set $I$. Moreover, the memory locations of the iterators generated by \tsmall{begin()} and \tsmall{end()} are $k_0$ and $k_0+(|I|+1)\cdot\delta$, respectively, where $\delta$ is the number of bytes to store an element as previously discussed. We have implemented two template classes \tsmall{fhg::iterator} and \tsmall{fhg::multi\_iterator} that allow to iterate over the multi-index set of a multidimensional array or a view. Objects of \tsmall{fhg::multi\_iterator} instantiate \tsmall{fhg::iterator} objects in order to define stride-based ranges. 

\subsection{Stride-based Iterator}
The iterator \tsmall{fhg::stride\_iterator} has the same traits as the iterator type of the \tsmall{std::vector} and therefore exhibits the same template class signature.
\begin{lstlisting}[classoffset=1,morekeywords={iterator_type,stride_iterator}]
template<class iterator_t>
class stride_iterator;
\end{lstlisting}
It provides all access properties of a random-access iterator and is therefore tagged as such. The template parameter \tsmall{iterator\_t} is an iterator type and must be accepted by the template class \tsmall{std::iterator\_traits} or by any of its pointer specializations for accessing the public member types such as \tsmall{iterator\_category}. We might think of the \tsmall{fhg::stride\_iterator} template class as an iterator adaptor for the standard random-access iterator with the same member functions and same interface. However, a stride-based iterator also features an additional member variable in order to store a stride of type \tsmall{std::size\_t}. The constructor has therefore the following signature.
\begin{lstlisting}[classoffset=1,morekeywords={iterator_t,stride_iterator}]
stride_iterator(iterator_t location, std::size_t stride);
\end{lstlisting}
An object of type \tsmall{iterator\_t} points to a valid memory location which we will denote as \tsmall{k}. The second parameter of the constructor determines the stride \tsmall{w} with which the iteration is performed. 
\begin{lstlisting}[classoffset=1,morekeywords={iterator_t,stride_iterator}]
stride_iterator& operator=(stride_iterator const& other);
\end{lstlisting}
The copy-assignment operator of the \tsmall{fhg::iterator} copies the current position \tsmall{k} and the stride \tsmall{w} of the \tsmall{other} argument. We therefore consider two dimension-based iterators \tsmall{i1} and \tsmall{i2} equal if the current positions \tsmall{i1.k}, \tsmall{i2.k} and the strides \tsmall{i1.w}, \tsmall{i2.w} of the iterators are equal.
% as it is depicted in the following listing.
%\begin{lstlisting}
%bool operator == (iterator const& i) const { return k == i.k && w == i.w; }
%bool operator != (iterator const& i) const { return k != i.k || w != i.w; }
%\end{lstlisting}
The expression \tsmall{(i1=i2) == i2} therefore returns true as both iterators have equal position and stride after the assignment \tsmall{(i1=i2)}. The following example illustrates how two stride-based iterators define a range for a given dimension. 
\begin{example}
\label{ex:stride.iterator.pair}
Let \tsmall{A} be a three dimensional array with elements of type \tsmall{float} stored according to the first-order storage format and let \tsmall{\{4,3,2\}} and \tsmall{\{1,2,3\}} be the shape and data layout tuple, respectively. The resulting stride tuple is then given by \tsmall{\{1,4,12\}} according to Eq.~\eqref{equ:stride.tuple.last.order}. We can then use the following two statements where the first line instantiates iterators of \tsmall{std::vector} and the second to iterators of the template class \tsmall{fhg::tensor}.
\begin{lstlisting}
std::vector<float>::iterator f1 {A.data()     }, l1 {A.data()+w[2]     };
fhg::tensor<float>::iterator f2 {A.data(),w[1]}, l2 {A.data()+w[2],w[1]};
\end{lstlisting}
The expressions \tsmall{A.data()} returns a pointer to first element. The array \tsmall{w} is the stride tuple of \tsmall{A}. The first half open range \tsmall{[f1,l1)} covers all elements with memory indices from \tsmall{0} and to \tsmall{12}. The second range \tsmall{[f2,l2)} only covers elements with the multi-indices $(0,i,0)$ for $0 \leq i < 2$ that corresponds to a fiber of the multidimensional array with elements that have the memory indices \tsmall{0}, \tsmall{4} and \tsmall{8}, see Eq.~\eqref{equ:f}. 
\end{example}
\begin{comment}
The user can also instantiate \tsmall{fhg::iterator} objects for a view of multidimensional arrays \tsmall{Av} using its offset and strides tuples that have been defined in Section~\eqref{sec:tensor:view}. The computation of the memory indices of a view is in that case accomplished using Eq.~\eqref{equ:lambda_view}. 
The following excerpt illustrates the implementation of some random-access iterator functions. The argument type \tsmall{ptrdiff\_t} is an integer type that can store the result of subtracting two pointers. 
\begin{lstlisting}[classoffset=1,morekeywords={ptrdiff_t}]
reference operator* (        ) const    {           return *k;   }
iterator& operator++(        )          { k +=   w; return *this;}
iterator& operator+=(std::ptrdiff_t n)  { k += n*w; return *this;}
\end{lstlisting}
%iterator  operator+ (diff_t n) const {           return iterator(k+n*w,w);}
\end{comment}
In order to iterate over the second dimension using the iterator pair \tsmall{[f1,l1)}, one has to explicitly provide the corresponding stride \tsmall{w[1]}. In case of the second iterator pair, no stride has to be specified within the loop.
\begin{example}
Let \tsmall{f1},\tsmall{l1} and \tsmall{f2},\tsmall{l2} be the iterator pair that have been defined in Example~\ref{ex:stride.iterator.pair}. Both of the following statements can be used to initialize the first row of the object \tsmall{A}.
\begin{lstlisting}
for(; f1 != l1;  f1+=w[1]) { *f1 = 5.0; }
for(; f2 != l2;  f2+=1   ) { *f2 = 5.0; }
\end{lstlisting}
\end{example}
Please note that the standard iterator \tsmall{f1} needs to be explicitly incremented with the stride of the array \tsmall{A}. The multidimensional iterator encapsulates the stride from the algorithm such that a normal increment with the \tsmall{operator++()} or \tsmall{operator+=(1)} suffices. Consider for instance the following function call of the \tsmall{std::fill} function that corresponds to the previous statements.
\begin{lstlisting}
std::fill(f2, l2, 5.0);
\end{lstlisting}
The iterator can therefore be used with all algorithms in the \tsmall{algorithm} and \tsmall{numeric} headers of the \tsmall{C++} standard library. For convenience, the user can also call member functions of the container classes to instantiate stride-based iterators and to define a range. 
\begin{lstlisting}
iterator begin(std::size_t dim);
iterator end  (std::size_t dim);
iterator begin(std::size_t dim, std::vector<std::size_t> const&);
iterator end  (std::size_t dim, std::vector<std::size_t> const&);
\end{lstlisting}
The first two functions define a range for a given dimension that must be smaller the order where the first element of the range always corresponds the first element of the corresponding container. The second argument of the last two functions correspond to a multi-index position that define the initial displacement within the multi-index space except the dimension that has been specified by \tsmall{dim}. The same initialization of the multidimensional container along the second dimension can be stated in one line.
\begin{lstlisting}
std::fill(A.begin(2),A.end(2),5.0);
\end{lstlisting}
The user does not have to keep track of the layout of the container \tsmall{A} and can read or write elements with the help of ranges that can be conveniently instantiated with appropriate function calls. Moreover, the user can combine fibers of multidimensional containers tuples using template functions of the \tsmall{C++} standard library. For instance, the inner product of two fibers of two multidimensional arrays with the same length can be simply expressed with the \tsmall{std::inner\_product} template function.
\begin{lstlisting}
std::inner_product(A.begin(3),A.end(3),B.begin(2),0.0);
\end{lstlisting}
Please note that the objects \tsmall{A} and \tsmall{B} can be instances of type \tsmall{tensor} or \tsmall{fhg::tensor\_view} with different layout tuples. We also provide \tsmall{begin()} and \tsmall{end()} member functions for the \tsmall{fhg::tensor} template class. The iterators have unit strides and define a range that covers the complete memory index set of the container class. They can be categorized as random-access iterators and behave just as iterators provided by \tsmall{std::vector}. The user can initialize all elements an \tsmall{fhg::tensor} object \tsmall{A} with the help of the \tsmall{std::fill} template function or the range-based for-loop and the range defined by \tsmall{A.begin()} and \tsmall{A.end()}. %The source file \tsmall{example2.cpp} contains utilization examples of algorithms from the \tsmall{C++} standard library .\todo{Korrigiere Beispiel.}  %We refer to 

\subsection{Multidimensional Iterator}
The template class \tsmall{fhg::multi\_iterator} defines a multidimensional iterator that allows to define the multi-index set of a multidimensional array or view. It is not used as an iterator per se, but functions as a factory class that instantiates stride-based iterators such as the previously described \tsmall{fhg::iterator} class. 
\begin{lstlisting}[classoffset=1,morekeywords={stride_iterator_t, multi_iterator}]
template<class stride_iterator_t>
class multi_iterator;
\end{lstlisting}
The template parameter \tsmall{stride\_iterator\_t} is an iterator type that must allow to instantiate stride-based iterators with a current position and a stride. The template class contains four private member variables:
\begin{itemize}
	\item the current pointer type \tsmall{pointer}, 
	\item the order of the container of type \tsmall{std::size\_t} and 
	\item two pointers to stride and extent tuples of type \tsmall{std::size\_t*}. 
\end{itemize}
The multidimensional iterator provides a single constructor with which all member variables are initialized.
\begin{lstlisting}
multi_iterator(pointer location,size_t rank,size_t* strides,size_t* extents);
\end{lstlisting}
The \tsmall{fhg::tensor} and \tsmall{fhg::tensor\_view} template classes offer the member functions \small{mbegin()} and \tsmall{mend()} to conveniently instantiate \tsmall{fhg::multi\_iterator} classes. Next to the default assignments, the template class \tsmall{fhg::multi\_iterator} overloads the assignment operator for stride-based iterators. 
\begin{lstlisting}[classoffset=1,morekeywords={stride_iterator_t, multi_iterator}]
multi_iterator& operator=(stride_iterator_t const& it);
\end{lstlisting}
The operator assigns the current pointer with the pointer of the argument. Next to the constructor, assignment and comparison operator, the \tsmall{fhg::multi\_iterator} provides two factory methods that instantiate stride-based iterators. 
\begin{lstlisting}[classoffset=1,morekeywords={stride_iterator_t, multi_iterator}]
stride_iterator_t begin(std::size_t r);
stride_iterator_t end  (std::size_t r);
\end{lstlisting}
The instantiated stride-based iterators are initialized with the dimension \tsmall{r} and the current position of the calling \tsmall{fhg::multi\_iterator}. In this way, we can iterate over the complete multi-index sets of multidimensional arrays and views with non-hierarchical data layouts. The following example illustrates how a three-dimensional array \tsmall{A} is iterated with the multidimensional and stride-based iterators \tsmall{fhg::multi\_iterator} and \tsmall{fhg::stride\_iterator}. Listing~\ref{alg:base} can be regarded as the baseline implementation for first-order template functions. 
\begin{lstlisting}[float,frame=single,frameround={tttt},numbers=left,numberfirstline=true,,caption=\footnotesize Baseline algorithm for higher-order tensor functions using multidimensional iterators, label=alg:base]
\end{lstlisting}
\begin{comment}
template <class InputIt, class Value>
void base(InputIt I, Value v) {
	for(auto j2 = I.begin(2), j2E = I.end(2); j2 != j2E; I=++j2)
		for(auto j1 = I.begin(1), j1E = I.end(1); j1 != j1E; I=++j1)
			for(auto j0 = I.begin(0), j0E = I.end(0); j0 != j0E; I=++j0)
				*j0 = v;
}
\end{comment}
Let \tsmall{A} be an object of type \tsmall{fhg::tensor} or \tsmall{fhg::tensor\_view} of order three storing its elements according to for instance the first-order storage format. Let \tsmall{I} be a multidimensional iterator that points to the first element of \tsmall{A}. We can then use the template function \tsmall{base} in Listing~\ref{alg:base} to initialize all elements with the value \tsmall{v}. The function generates a range with the stride-based iterator \tsmall{j2} and \tsmall{j2End} for the outermost dimension. Each increment operation \tsmall{++j2} adds its stride to its internal pointer until the end of the range is reached. Using the assignment operator of \tsmall{I}, the internal pointer of the stride-based iterator \tsmall{j2} is copied to \tsmall{I} with which the stride-based iterator \tsmall{j1} is initialized. The stride-based iterator \tsmall{j0} is initialized in the same manner. Finally, the elements of \tsmall{A} are set to the value \tsmall{v}. We can also replace the innermost loop with 
\begin{lstlisting}
std::fill(I.begin(0), I.end(0), v);
\end{lstlisting}
where \tsmall{std::fill} is the template function provided by the \tsmall{C++} standard library.
%TODO: Similar iterators have been also proposed in \todo{Abgrenzung zu anderen Arbeiten.} 

\section{Tensor Function Templates}
\label{sec:functions}
Our tensor algorithms are implemented with multidimensional iterators and support a combination of multidimensional arrays and views with different data layouts, arbitrary order and extents. All of following tensor algorithms are parametrized by multidimensional iterator types that need
\begin{itemize}
	\item to provide two member functions \tsmall{begin()} and \tsmall{end()} in order to select and iterate over the corresponding index range and
	\item to have the same functionality as a standard iterator with at least the input iterator capabilities.
\end{itemize}
Additional requirements of a dimension specific iteration depends on the algorithm that may require different iterator capabilities. An example of such an iterator is the multidimensional iterator template class \tsmall{fhg::multi\_iterator} that has been introduced in the previous section. Users can specify their own iterator type and use our tensor template functions. Please note that the member functions of \tsmall{fhg::tensor} and \tsmall{fhg::tensor\_view} call the following first- and if applicable higher-order functions. Both data structures, for instance, overload the relational and numeric operators. They provide a convenient interface that allows to express numerical algorithms close to the mathematical notation without specifying multidimensional iterators. The user may decide whether to call the higher-order tensor functions using member functions of tensor template class instances or to call them with a different multidimensional container or multidimensional iterator. 
\subsection{First-Level Tensor Operations}
\label{sec:function.templates.scalar}
\begin{table}[t]
	%\scriptsize
	%\small
	\begin{center}
		\footnotesize
		\begin{tabular}{lll}
			\toprule
			Function & Example (\texttt{MATLAB}) &  Description\\%Function interface\\
			\midrule
			
			\texttt{for\_each()}  & \texttt{C = C+3} & Performs a unary operation for each element \\	
			
			\texttt{copy()}       & \texttt{C = A} & Copies elements starting with the first element \\
			
			\texttt{copy\_if()}   & \texttt{C(A>3) = A(A>3)} & Copies elements that match a criterion \\
			
			\texttt{transform()}  & \texttt{C = A+3} & Modifies, copies elements accord. to a unary op. \\
			
			\texttt{transform()}  & \texttt{C = A+B} & Combines elements accord. to a binary op. \\
			
			\texttt{fill()}       & \texttt{C = 3} & Replaces each element with a given value \\
			
			\texttt{generate()}   &  \texttt{C(i) = i.*i} & Replaces each element with the result of an op. \\

			\midrule
			
			\texttt{count()}        & \texttt{sum(A(:)==3)} & Returns the number of elements \\	
			\texttt{count\_if()}    & \texttt{sum(A(:)<=2)} & Returns the number of elements matching a crit. \\

			\texttt{min\_element()} & \texttt{[,i] = min(A(:))} & Returns the element with the smallest value \\
			\texttt{max\_element()} & \texttt{[,i] = max(A(:))} & Returns the element with the largest value \\
			
			\texttt{find()}         & \texttt{find(C(:)==3)} & Searches for the first element matching the value \\
			\texttt{find\_if()}     & \texttt{find(C(:)<=2)} & Searches for the first element matching a crit. \\
			
			\texttt{equal()}        & \texttt{all(C(:)==A(:)} & Returns whether two ranges are equal \\
			\texttt{mismatch()}     & \texttt{find(C(:)!=A(:)} & Returns the first elements that differ \\
			
			\texttt{all\_of()}      & \texttt{all(C(:)==3)}    &  Returns whether all elements match a crit. \\
			\texttt{any\_of()}      & \texttt{sum(C(:)==3)>0}  &  Returns whether at least one element matches a crit. \\
			\texttt{none\_of()}     & \texttt{sum(C(:)==3)==0} &  Returns whether none of the elements matches a crit. \\
			
			\midrule
			
			\texttt{iota()}         &  \texttt{C(:)} \texttt{= [1:n]} & Replaces each element with incremented values \\ 
			\texttt{accumulate()}   &  \texttt{sum(C(:))} & Combines all element values (accord. to a binary op.) \\ 
			\texttt{inner\_product()} & \texttt{dot(C(:),A(:))} & Combines all elements (accord. to a binary op.)\\ 
			\bottomrule
		\end{tabular}
	\end{center}
	\caption{\footnotesize Overview of the implemented higher-order tensor functions where the algorithms can combine multidimensional arrays with different storage layouts. The symbol $\odot$ denotes a binary operator for real numbers and implements e.g. a multiplication or addition. The transform, compare as well as the tensor-multiplication algorithms allow to transpose the multidimensional arrays according to the permutation tuples before applying the corresponding operation.}
	\label{tab:pointwise_operations}
\end{table}%

%%%%%%%%%%%%%%%%%%%%%%%%%%%

% Differences to the \tsmall{C++} standard algorithms
First-level tensor template functions such as \tsmall{fhg::for\_each} implement algorithms of the \tsmall{C++} standard library for dense multidimensional arrays and their views. Table~\ref{tab:pointwise_operations} lists some of the implemented template functions of our tensor framework that make use of multidimensional iterators. 
Similar to the \tsmall{C++} standard library, the user can pass function objects to some of the first-level template functions that apply one or more function objects in each iteration. Functions objects are function-like objects that provide a function call operator allowing the function to have a state. The user can define its own class with a function call operator to instantiate a function object or construct a function object using lambda-expressions. Most of our first-level template functions require unary or binary predicates and unary or binary function objects just as it is the case for the \tsmall{C++} standard library.

In contrast to the \tsmall{C++} standard algorithms, our first-level tensor functions are designed for multidimensional arrays and views. They are implemented in a recursive fashion and iterate over multiple half-open ranges. The iteration is accomplished with a multidimensional iterator where one or more elements of multidimensional arrays and views with different data layouts but the same multi-index are combined. Please note that dense multidimensional arrays with a contiguous data layout can also be manipulated with template functions of the \tsmall{C++} standard library. However, they are designed for containers with one dimension and therefore cannot manipulate arrays with different data layouts or with non-contiguously stored elements. 

% Discuss how we can derive the recursive template class functions 
Please consider the \tsmall{fhg::for\_each} template function in Listing~\ref{alg:foreach}. The template function is recursively iterates over the complete memory index set of a multidimensional array using multidimensional iterators of type \tsmall{InputIt} and applies the unary function of type \tsmall{UnaryFn} to the elements that are referenced by the iterator. The implementation accepts any order greater zero with which \tsmall{fhg::for\_each} is called and therefore is the generalized version of the \tsmall{base} function that has been provided in the previous section in Listing~\ref{alg:base}. The iteration of the memory index set is accomplished by recursively adding memory indices in each function call with a newly defined index range given by \tsmall{in.begin(r)} and \tsmall{in.end(r)}. 
%Please note, that the only difference of the function signature is the missing \tsmall{last} argument for definition of an half-open range. 
%Please note again, that a single half-open range does not suffice in order to iterate along a multidimensional index set without imposing restrictions on the underlying data type. 
\begin{lstlisting}[float,frame=single,frameround={tttt},numbers=left,numberfirstline=true,caption=\footnotesize Template function \tsmall{fhg::for\_each} based on multidimensional iterators,label=alg:foreach, classoffset=1,morekeywords={SizeT, InputIt, UnaryFn}]
\end{lstlisting}
\begin{comment}
template <class InputIt, class UnaryFn> 
void for_each(std::size_t r, InputIt in, UnaryFn fn){ 
	auto f = in.begin(r), l = in.end(r);
	if(r > 0)
		for(; f != l; in=++f)
			for_each(r-1,in,fn) 
	else
		std::for_each(f,l,fn);
} 
\end{comment}
More specifically, if \tsmall{p} is the order of the multidimensional array, the function \tsmall{for\_each} needs to be initially called with \tsmall{p-1}. In this case, the outermost loop iterates over the half-open range that is defined by \tsmall{it.begin(p-1)} and \tsmall{it.end(p-1)} covering the index range of the $r$-th dimension. The copy-constructed iterator \tsmall{f} is then copied to the next function instance where $r=p-2$. This is repeated until \tsmall{r} equals zero where the first dimension of the data structure is accessed and the \tsmall{std::for\_each} template function of the \tsmall{C++} standard library is called. The multidimensional iterators \tsmall{f} and \tsmall{l} define a range and act like a standard iterator with input iterator capabilities. The maximum depth of the recursion is therefore \tsmall{p-1} and the recursive function is called $n_2 \cdots n_p$ times where $n_r$ is the $r$-th extent of the multidimensional array or view. %The time complexity of the \tsmall{fhg::for\_each} function is therefore $\mathcal{O}(n_1 \cdots n_p)$.

Incrementing the $r$-th stride-based iterator in the $r$-th loop corresponds to a shift of its internal pointer by the $r$-th stride $w_r$. The $i$-th iteration therefore sets the internal pointer of the $r$-th stride-based iterator to $k$ with $k = k' + \delta \cdot i_r \cdot w_r$, where $k'$ is the memory address computed by the previous stride-based iterator and $\delta$ is the number of bytes to store an element of \tsmall{A}. Given zero-based indices $i_r$ for all $r$, the address $j$ of an element is given by 
\begin{equation}
\label{equ:final_address}
\begin{split}
j &= k_0 + \delta \cdot \sum_{r=1}^p i_r \cdot w_r = k_0 + \delta \cdot \lambda(\mbw,\mbi,\mathbf{0}),
\end{split}
\end{equation}
where $\lambda$ is the layout function already defined in Eq.~\eqref{equ:absolute.memory.address}. In case of a multidimensional array view, the $r$-th stride of the multidimensional iterator is given by $w_r\cdot t_r$ where $t_r$ is step of the $r$-th range and $w_r$ is the stride of the referenced multidimensional array as discussed in Section~\ref{sec:tensor:view}. Additionally, the index of the first element is initialized with $w_1\cdot f_1 + \cdots + w_p\cdot f_p$ where $f_r$ is the $r$-th lower bound of the view that has been used in Eq.~\eqref{equ:lambda_view}. 
\begin{example}
Let $\mubA$ be a tensor or view of order $p$ with the shape tuple $\mbn$ and any non-hierarchical layout. The following equation denotes the initialization of $\mubA$ with a scalar.
\begin{equation}
\label{equ:initialize}
\mubA(i_1,\dots,i_p) = \alpha, \quad \text{for all} \ i_r \in I_r.
\end{equation}
Given a tensor object \tsmall{A} with elements of type \tsmall{float}, the next listing implements Eq.~\eqref{equ:initialize} and initializes elements of an array \tsmall{A} with the value \tsmall{alpha} using the template function \tsmall{fhg::for\_each}.
\begin{lstlisting} 
fhg::for_each(A.mbegin(), A.mend(), [alpha](float& a){a=alpha;}); 
\end{lstlisting} 
Note that the object \tsmall{A} can be a view and may have any arbitrary non-hierarchical layout.
\end{example}
Please note the similarity of the function signature of \tsmall{fhg::for\_each} with the one of \tsmall{std::for\_each}. In case of \tsmall{fhg::for\_each}, the first argument must be a multidimensional iterator that is capable of creating multiple half-open index ranges. The second argument must be a unary function that is accepted, i.e. applicable by the \tsmall{std::for\_each} function. The first template parameter of the function indicates that the multidimensional iterator needs only to have input iterator traits. This might be different for other template functions such as \tsmall{fhg::sort} postulating a random iterator.

Consider the \tsmall{fhg::transform} template function in Listing~\ref{alg:transform}. The implementation recursively applies the unary function \tsmall{fn} to elements of a multidimensional array that is referenced by the multidimensional input iterator \tsmall{in}. The results are stored in an array with the same order and extents using the multidimensional output iterator \tsmall{out}. The template functions returns an iterator to the element past the last transformed element. The first iterator \tsmall{in} points to and iterates over a multidimensional array or view with a data layout that may be different from the multidimensional array or view referenced by the second iterator. 
\begin{lstlisting}[float,frame=single,frameround={tttt},numbers=left,numberfirstline=true,,caption=\footnotesize Template function \tsmall{fhg::transform} based on multidimensional iterators, label=alg:transform, classoffset=1,morekeywords={InputIt, OutputIt, UnaryFn}]
\end{lstlisting}
\begin{comment}
template <class InputIt, class OutputIt, class UnaryFn> 
OutputIt transform(std::size_t r, InputIt in, OutputIt out, UnaryFn fn)
{
	auto fin  = in .begin(r), lin = in.end(r); 
	auto fout = out.begin(r); 
	if(r > 0) 
		for(; fin!=lin; in=++fin, out=++fout) 
			out = transform(r-1, in, out, fn);
	else
		out = std::transform (fin, lin, fout, fn); 
	return out;
} 
\end{comment}
The control flow is similar to the one of the \tsmall{fhg::for\_each} function with the difference that each function call returns the current output pointer. The innermost loop calls the \tsmall{C++} standard library \tsmall{std::transform} template function. The dereferenced input and output iterators correspond to elements with the same multi-index regardless of the data layout. All template functions listed in Table~\ref{tab:pointwise_operations} are implemented in the same manner. 
\begin{example}
Let $\mubA$ and $\mubC$ be multidimensional arrays of order $p$ with the same shape tuple $\mbn$. The multiplication of a scalar $\alpha$ with elements of a tensor $\mubA$ is then given by the equation
\begin{equation}
\label{equ:scaling}
\mubC(i_1,\dots,i_p) = \alpha \cdot \mubA(i_1,\dots,i_p), 
\end{equation}
where $i_r \in I_r$ and $r = 1,\dots,p$. Let \tsmall{A} and \tsmall{C} be objects of type \tsmall{fhg::tensor} or \tsmall{fhg::tensor\_view} with the same shape tuple and let \tsmall{alpha} be a scalar. Eq.~\eqref{equ:scaling} can then be implemented with help of the \tsmall{fhg::transform} function.
\begin{lstlisting} 
fhg::transform(A.mbegin(), C.mbegin(), [alpha](float const& a){return a*alpha;}); 
\end{lstlisting}
\end{example}
\begin{example}
Let $\mubA$, $\mubB$ and $\mubC$ be multidimensional arrays with same shape tuples $\mbn$ and $\odot$ a binary operation that is either an addition, subtraction, multiplication or division. The elements of $\mubC$ are then given by the equation
\begin{equation}
\label{equ:transform}
\mubC(i_1,\dots,i_p) = \mubA(i_1,\dots,i_p) \odot \mubB(i_1,\dots,i_p),
\end{equation}
where $i_r \in I_r$ and $r = 1,\dots,p$. Let \tsmall{A}, \tsmall{B} and \tsmall{C} be objects of type \tsmall{fhg::tensor} or \tsmall{fhg::tensor\_view} with the same shape tuple. Similar to the previous example, Eq.~\eqref{equ:transform} can be implemented using the \tsmall{fhg::transform} function.
\begin{lstlisting} 
fhg::transform(A.mbegin(), B.mbegin(), C.mbegin(), std::plus<>()); 
\end{lstlisting}
\end{example}
Please note that we did not specify the storage formats and index offsets of the tensor data types. User of the framework is free to implement different types operations with appropriate binary operators using their own template classes or multidimensional iterators. 
%We have used the same template functions for the implementation of the arithmetic operators for the \tsmall{fhg::tensor} and \tsmall{fhg::tensor\_view} template classes which allows us to write code like \tsmall{if(A==B)} \tsmall{A+=B*3;}. 
For convenience, the \tsmall{fhg::tensor} and \tsmall{fhg::tensor\_view} template classes provide member functions that perform entrywise operations with overloaded operators for different argument types. The overloaded \tsmall{operator+=()} member function of both template classes calls the \tsmall{fhg::transform} template function if its arguments are of type \tsmall{fhg::tensor} and \tsmall{fhg::tensor\_view}. If the argument is a scalar of type \tsmall{const\_reference} or \tsmall{value\_type}, the template function \tsmall{fhg::for\_each} is called, see Example~\ref{equ:initialize}. 
%We refer to the \tsmall{example4.cpp} that illustrate the usage of first-level tensor operations.\todo{Change the example3.cpp}

\subsection{Higher-Level Tensor Operations}
\label{sec:function.templates.multiplication}
Higher-level tensor operations exhibit a higher arithmetic intensity ratio than first-level tensor operations and perform one or more inner products over specified dimensions. %For instance, the \tsmall{k}-mode tensor-vector multiplication of a multidimensional array of order \tsmall{p} with a vector consisting of \tsmall{nk} elements, involves the computation of \tsmall{n1*n2*...*np} inner products where the vector is accessed multiple times. 
Table~\ref{tab:higher-order-operations} provides an overview of the implemented higher-order tensor operations. We will discuss the implementation of the three higher-order tensor operations, that are the tensor-times-vector and tensor-times matrix and tensor-times-tensor multiplication. %
%\todo[noline]{Need to include the transposition of tensors for the operation.}
Please note that the transposition is included in this subsection as it can be expressed with a copy operation and a modified stride tuple.
\begin{table}[t]
	%\scriptsize
	%\small
	\begin{center}
		\small
		\begin{tabular}{lll}
			\toprule
			Function & Notation & Example \\%Function interface\\
			\midrule
			\tsmall{transpose()}              & $\mubC \leftarrow \mubA^{\mbtau} $                                  & \tsmall{C = permute(A,[2,1,3])} \\	
			\tsmall{tensor\_times\_vector()}  & $\mubC \leftarrow \mubA \bar{\bullet}_m \mbb $                      & \tsmall{C = ttv(A,b,1)} \\
			\tsmall{tensor\_times\_matrix()}  & $\mubC \leftarrow \mubA \bullet_m \mbB$                             & \tsmall{C = ttm(A,B,1)} \\
			\tsmall{tensor\_times\_tensor()}  & $\mubC \leftarrow \mubA \bullet_q^{\varphi,\psi} \mubB $            & \tsmall{C = ttt(A,B,[1 3],[2 3])} \\
			\tsmall{outer\_product()}         & $\mubC \leftarrow \mubA \circ \mubB $                               & \tsmall{C = ttt(A,B)} \\
			\tsmall{inner\_product()}         & $\alpha \leftarrow \langle \mubA,\mubB \rangle$                     & \tsmall{C = ttt(A,B,[1:3])} \\
			\tsmall{norm()}                   & $\alpha \leftarrow | \mubA |_F$                                     & \tsmall{C = norm(A)} \\
			\midrule
			\tsmall{tensor\_times\_vectors()}  & $\mubC \leftarrow \mubA \bar{\bullet}_r \mbv \bar{\bullet}_q \mbu$ & \tsmall{C = ttv(A,u,v,[1,3])} \\
			\tsmall{tensor\_times\_matrices()} & $\mubC \leftarrow \mubA \bullet_r \mbU \bullet_q \mbV$             & \tsmall{C = ttm(A,U,V,[1,3])} \\
			\bottomrule
		\end{tabular}
	\end{center}
	\caption{Overview of the implemented higher-order tensor functions. The first column lists template function  that implement the corresponding operations listed in the second column. The third column provides example code that can be run in \tsmall{Matlab} using the toolbox presented in~\cite{bader:2006:algorithm862}.}
	\label{tab:higher-order-operations}
\end{table}%
\begin{comment}
Let $\mubA$ be a multidimensional array or order \tsmall{p} with the shape tuple $\mbn_a$. Let $\mubC$ be another multidimensional array with a shape tuple $\mbn_c$. The transposition of the multidimensional array $\mubA$ with the \textit{transposition tuple} $\mbtau$ is given by
\begin{equation}
\label{equ:transposition}
\mubC(i^c_1,\dots,i^c_p) = \mubA(i^a_1,\dots,i^a_p) \quad \text{with} \quad i^c_r = i^a_{\tau_r},
\end{equation}
where the shape tuples satisfy $\mbn_c[r]= \mbn_a[\tau_r]$ for all $r$ dimensions.
\begin{lstlisting}[float,frame=single,frameround={tttt},numbers=left,numberfirstline=true,,caption=\footnotesize Template Function \tsmall{transpose} using Multidimensional Iterators implementing Eq.~\eqref{equ:transposition}, label=alg:transposition, classoffset=1,morekeywords={InputIt1,OutputIt}]
template <class InputIt, class OutputIt>
void transpose(size_t r, size_t* tau, InputIt in, OutputIt out)
{
	auto fin  = in.begin(tau[r]-1), lin = in.end (tau[r]-1);
	auto fout = out.begin(r);
	
	if(r > 0)
		for(; fin != lin; ++fin, ++fout)
			fhg::detail::transpose(r-1, tau, in=fin, out=fout);
	else
		std::copy(fin,lin,fout);
}
\end{lstlisting}
\end{comment}

Let $\mubA$ be an $p$-way array with the shape tuple $\mbn$ and let $\mbb$ be a $1$-way array with the shape tuple $(n_m)$. Let also $\mubC$ be a $p-1$-way array with its shape tuple $\mbn = (n_1,\dots,n_{m-1},n_{m+1},\dots,n_p)$. The \textit{$m$-mode tensor-vector multiplication}, denoted by $\mubC = \mubA \bar{\bullet}_m \mbb$, multiplies $\mubA$ with $\mbb$ according to 
\begin{equation}
\label{equ:ttv}
\mubC(i_1,\dots,i_{m-1},i_m,\dots,i_{p}) = \sum_{i_m=0}^{n_m-1} \mubA(i_1,\dots,i_p) \cdot \mbb(i_m),
\end{equation}
where $1 \leq m \leq p$ and $2 \leq p$. Please note that the tensor-vector product equals to a common vector-matrix multiplication if $m=1$ and $p=2$. For $p > 2$, the vector $\mbb$ is multiplied with each frontal slice of the array $\mubA$. Listing~\ref{alg:ttv} implements the tensor-times-vector multiplication according to Eq.~\eqref{equ:ttv} where the contracting dimension $m$ must be greater than zero. Our framework provides also an implementation for this case. The second and third arguments, $r$ and $q$, track the recursion depth where $q=r$ for $m<r<p$ and $q-1=r$ for $0<r\leq m$ and are initially set to $r=p-1$ and $q=p-1$.
\begin{lstlisting}[float,frame=single,frameround={tttt},numbers=left,numberfirstline=true,,caption=\footnotesize Template Function \tsmall{ttv} using Multidimensional Iterators implementing Eq.~\eqref{equ:ttv}, label=alg:ttv, classoffset=1,morekeywords={Unsigned,InputIt1,InputIt2,OutputIt,multiplies}]
\end{lstlisting}

\begin{comment}
template <class Unsigned, class InputIt1, class InputIt2, class OutputIt>
void ttv(Unsigned m,Unsigned r,Unsigned q, InputIt1 A, InputIt2 B, OutputIt C)
{
	if(r == m) {
		ttv(m,r-1 q, A, B, C);
	}
	else if(r == 0){
		auto fa = A.begin(0), la = A.end(0);
		for(auto fc = C.begin(0); fa != la; A=++fa, ++fc)
			 *fc = std::inner_product(A.begin(m), A.end(m), B.begin(), *fc, 
			              std::plus<>(), std::multiplies<>());
	}
	else {
		auto fa = A.begin(r), la = A.end(r); 
		for(auto fc = C.begin(q); fa != la; A=++fa, C=++fc) {
			ttv(m, r-1, q-1, A, B, C);
	}	
}
\end{comment}

The \tsmall{fhg::ttv} template function has a similar control flow compared to the previously discussed implementation of the \tsmall{fhg::transform} template function. One difference is the recursive call in line 5 where the recursion depth $r$ equals the mode $m$. After $p-1$ recursive function calls with $r=0$, the inner product of two fibers is performed in line 10 using the \tsmall{std::inner\_product} of the standard library. The range of the first fiber is given by the range that is defined by the iterators \tsmall{A.begin(m)} and \tsmall{A.end(m)}. The start of the second range is given by \tsmall{B.begin(0)}. The result is stored into the location pointed by \tsmall{fc}. Interpreting lines 8 to 11 as slice-vector multiplications, we might say that the tensor-vector multiplication is composed of multiple slice-times-vector multiplications. 
\begin{example}
Let\tsmall{A}, \tsmall{b} and \tsmall{C} are tensor objects of the same element type and layout and let the shape tuples of \tsmall{A}, \tsmall{b} and \tsmall{C} be given by the tuples $(3,4,2)$, $(2,1)$ and $(3,4)$, respectively. We can then use  following statement that performs the $2$-mode tensor-times-vector multiplication.
\begin{lstlisting}
fhg::ttv(1, 2, 2, A.mbegin(), b.mbegin(), C.mbegin());
\end{lstlisting}
\end{example}
A second implementation with only two branches is given when $m=0$. The first branch contains the expressions of the \tsmall{else}-branch that are executed for $r>1$. The second branch contains the computation of a loop over the second dimension (instead of the first) and computes the inner product of two fibers of the first dimension. We might therefore say that for $m=0$, the tensor-times-vector multiplication consists of multiple vector-times-matrix multiplication where the matrix is the frontal slice of the first input array pointed by \tsmall{A}.

Let $\mubA$ be an $p$-way array with the shape tuple $\mbn$ and let $\mbB$ be a $2$-way array with the shape tuple $(n',n_m)$. Let also $\mubA$ be an $p$-way array with the shape tuple $(n_1,\dots,n_{m-1},n',n_{m+1},\dots,n_p)$. The \textit{$m$-mode tensor-matrix multiplication}, denoted by $\mubC = \mubA \bullet_m \mbB$, multiplies $\mubA$ with $\mbB$ according to 
\begin{equation}
\label{equ:ttm}
\mubC(i_1,\dots,i_{m-1},j,i_{m+1},\dots,i_p) = \sum_{i_m=0}^{n_m-1} \mubA(i_1,\dots,i_p) \cdot \mbB(j,i_m),
\end{equation}
with $1 \leq m \leq p$ and $2 \leq p$. The tensor-matrix multiplication corresponds to a tensor-vector product when $n'=1$. If the order $p$ of the arrays $\mubA$ and $\mubC$ equals $2$, the tensor-matrix product corresponds to a common matrix-matrix multiplication. 

The implementation of tensor-matrix multiplication is almost identical to the recursive template function \tsmall{ttv} where only the control flow of the most inner recursion differs. Moreover, the tensor-times-matrix multiplication can be expressed in terms of multiple slice-matrix multiplications where one parameter suffices in order to generate ranges for the input and output iterators. Consider Listing~\ref{alg:ttm} that illustrates the implementation of the innermost part of the template function \tsmall{fhg::ttm}. It is the base case of the recursion and corresponds to a slice-times-matrix multiplication. Parameter \tsmall{m} is the contraction dimension and \tsmall{r} denotes the recursion depth that initially is set to $p-1$. Please note that $m$ needs to be greater than zero in this case. Our framework provides a second implementation for $m=0$ with a control and data flow that is similar to the tensor-times-vector implementation.
\begin{lstlisting}[float,frame=single,frameround={tttt},numbers=left,numberfirstline=true,caption=\footnotesize Most inner Recursion Level of the Template Function \tsmall{ttm}, label=alg:ttm, classoffset=1,morekeywords={Unsigned,InputIt1,InputIt2,OutputIt,multiplies}]
\end{lstlisting}

\begin{comment}
template <class Unsigned, class InputIt1, class InputIt2, class OutputIt>
void ttm(Unsigned m, Unsigned r, InputIt1 A, InputIt2 B, OutputIt C)
{
	//...
	auto fa0 = A.begin(0), la0 = A.end(0);
	auto fb0 = B.begin(0);
	for(auto fc0 = C.begin(0); fa0 != la0; A=++fa0, C=++fc0, B=fb0) {
		auto fcm = C.begin(m), lcm = C.end(m);
		for(auto fb = fb0; fcm != lcm; C=++fcm, B=++fb)
			*fcm = std::inner_product( A.begin(m), A.end(m), B.begin(1), *fcm, 
			                                std::plus<>(), std::multiplies<>());
	}
	//...
}
\end{comment}

The initial addresses of the input and output slices are generated in line 5 and 7 by calling \tsmall{A.begin(0)} and \tsmall{C.begin(0)}, respectively. The initial address of the input matrix is generated in line 6 by calling \tsmall{B.begin(0)}. The instantiated stride-based iterators \tsmall{fa0}, \tsmall{fb0} and \tsmall{fc0} iterate in lines 7 and 9 along the first dimension of the corresponding data structures and point to fibers and rows of the input and output slices, respectively. The range of the output fiber is determined in line 8 by the instantiation of the stride based iterators \tsmall{fcm} and \tsmall{lcm}. The value for each element of the output fiber is given by the inner product of the input fiber and input matrix row in lines 10 and 11. The range of the input fiber and matrix row are given by the ranges (\tsmall{A.begin(m)},\tsmall{A.end(m)}) and (\tsmall{B.begin(1)}, \tsmall{B.end(1)}), respectively. Please note that iterator \tsmall{B} needs to be initialized with the initial address in line 7. %In summary, lines 8 to 11 are equivalent to a fiber-times-matrix multiplication. We therefore can say that the tensor-times-matrix multiplication consists of multiple fiber-times-matrix multiplications. 

The tensor-tensor product is the general form of the tensor-matrix and tensor-vector multiplication. Let $\mubA$ be $(q+r)$-way array with $\mbn_a = (n^a_1,\dots,n^a_{q+r})$ and let $\mubB$ be a $(q+s)$-way array with $\mbn_b = (n^b_1,\dots, n^b_{q+s})$ where $r,s,q \in \bbN_0$ and $r + q > 0$, $s + q > 0$. The \textit{$(q,\varphi,\psi)$-mode tensor-tensor multiplication}, depicted by $\mubA \bullet_{q}^{\varphi,\psi} \mubB$, computes elements of the $(r+s)$-way array $\mubC$ according to
\begin{equation}
\label{equ:ttt}
%\begin{split}
\mubC(i^c_1,\dots,i^c_{r+s}) =
%&\qquad \qquad 
\sum_{j_1=0}^{n_1'-1} \cdots \sum_{j_{q}=0}^{n_{q}'-1} \mubA(i^a_1,\dots,i^a_{q+r}) \cdot \mubB(i^b_1,\dots,i^b_{q+s}), 
%\end{split}
\end{equation}
where the shape tuples satisfy the following equations:
\begin{equation}
\label{equ:ttt_indices}
\begin{alignedat}{3}
n^c_k                      &= n^a_{\varphi_k}   &\quad &\text{for} \ 1\leq k \leq r, \\
n^c_{k+r}                  &= n^b_{\psi_k}      &\quad &\text{for} \ 1\leq k \leq s, \\
n_k' = n^a_{\varphi_{k+r}} &= n^b_{\psi_{k+s}}  &\quad &\text{for} \ 1\leq k \leq q.\\
%
%n_{k+r}^c                    &= n^b_{\varphi^b_k}        &\quad &\text{for} \ 1\leq k \leq s, \\
%n_k' = n^a_{\varphi^a_{k+r}} &= n^b_{\varphi^b_{k+s}}  &\quad &\text{for} \ 1\leq k \leq q.\\
\end{alignedat}
\end{equation}
The first $r$,$s$ elements of the permutation tuples $\mbvarphi_a$ and $\mbvarphi_b$, respectively, determine the dimension ordering for $\mubC$. The last $q$ elements determine the dimensions that are contracted. 

Equations~\eqref{equ:ttt} and~\eqref{equ:ttt_indices} can be used to define the tensor-matrix and tensor-vector multiplication. Let $\mbn_a = (n_1^a,\dots,n_p^a)$ and $\mbn_b = (n_m^a)$ be the shape tuples of the first and second operand. A tensor-tensor multiplication is the $m$-mode tensor-vector multiplication, if $q=1$, $r = p-1$, $s = 0$ and $\mbvarphi = (1,\dots,m-1,m+1,\dots,p,m)$, $\mbpsi = (1)$ such that $\mbn_c = (n^a_1,\dots,n^a_{m-1}, n^a_{m+1}, \dots, n^a_{p})$. The $m$-mode tensor-matrix multiplication is given if $m=p$ and $n^b_2 = n_p^a$, where $q=1$, $r = p-1$, $s = 1$ and $\mbvarphi = (1,\dots,p-1,p)$, $\mbpsi = (1,2)$ such that $\mbn_c = (n^a_1,\dots,n^a_{p-1},n^b_{1})$. The inner product is given when $r=0$ and $s=0$ for some $q>0$, while the outer product is given when $q=0$ for some $r>0$ and $s>0$.

Listing~\ref{alg:ttt} illustrates the recursive implementation of the tensor-times-tensor multiplication as defined in Eq.~\eqref{equ:ttt}. The implementation is recursive and performs the contraction without unfolding the tensors using multidimensional iterators only. Let us first consider the function signature of \tsmall{fhg::ttt}. 
%\begin{lstlisting}[classoffset=1,morekeywords={Unsigned, Permutation, InputIt1,InputIt2,OutputIt}]
%template <class Unsigned, class Permutation, 
%          class InputIt1, class InputIt2, class OutputIt>
%void ttt(Unsigned k, Unsigned q, Unsigned r, Unsigned s, 
%         Permutation phi, Permutation psi, InputIt1 A, InputIt2 B, OutputIt C)
%\end{lstlisting}
The first parameter \tsmall{k} is the recursion depth of the template function. The parameter \tsmall{q} is the number of dimensions of the input arrays that are contracted. The parameters \tsmall{r} and \tsmall{s} are the remaining number of dimensions of the input arrays that sum up to the order of the output array. The next two variables \tsmall{phi} and \tsmall{psi} are or point to permutation tuples that are needed to specify the dimension ordering of the output array and the contraction. The template function expects all integer variables to be zero-based. The control flow of the recursive function \tsmall{fhg::ttt} contains four branches using \tsmall{if}-\tsmall{else} statements. In each branch multidimensional iterators are adjusted and stride-based iterators are instantiated and incremented with respect to the equations~\eqref{equ:ttt} and~\eqref{equ:ttt_indices}. The \tsmall{else} statement performs the innermost loop of the tensor multiplication which computes the inner product of two fibers.
\begin{lstlisting}[float,frame=single,frameround={tttt},numbers=left,numberfirstline=true,caption=\footnotesize Template Function \tsmall{ttt} using multidimensional iterators implementing Eq.~\eqref{equ:ttt}, label=alg:ttt, classoffset=1,morekeywords={Unsigned, Permutation, InputIt1,InputIt2,OutputIt,multiplies}]
\end{lstlisting}
\begin{comment}
template <class Unsigned, class Permutation, 
          class InputIt1, class InputIt2, class OutputIt>
void ttt(Unsigned k, Unsigned q, Unsigned r, Unsigned s, 
         Permutation phi, Permutation psi, InputIt1 A, InputIt2 B, OutputIt C)
{
    if(k < r) {
	    auto fa = A.begin(phi[k]), la = A.end(phi[k]);
	    for(auto fc = C.begin(k); fa!=la; A=++fa,C=++fc) {ttt(k+1,...);}
    }
    else if(k < r+s) {    
	    auto fb = B.begin(psi[k-r]), lb = B.end(psi[k-r]);
	    for(auto fc = C.begin(k); fb!=lb; B=++fb,C=++fc) {ttt(k+1,...);}
    }
    else if(k < r+s+q-1) {
	    auto fb = B.begin(psi[k-r]), lb = B.end(psi[k-r]);
	    for(auto fa = A.begin(phi[k-s]-1); fb!=lb; B=++fb,A=++fa) {ttt(k+1,...);}
    }
    else {
	    *C = std::inner_product(A.begin(phi[k-s]), A.end(phi[k-s]), 
	          B.begin(psi[k-r]-1), *C, std::plus<>{}, std::multiplies<>{});
    }
}
\end{comment}

The recursive template function \tsmall{fhg::ttt} is called initially with $k=0$. For $k<r$, the control flow branches into the loop in line 7 that calls the function \tsmall{fhg::ttt} with modified iterators \tsmall{A}, \tsmall{C} and with \tsmall{k+1}. The iteration length of the $k$-th loop where $k$ is smaller than $r$ is determined by the stride-based iterators \tsmall{fa}, \tsmall{la} and \tsmall{fc} which are instantiated with the multidimensional input and output iterators \tsmall{A} and \tsmall{C}, respectively. The iterator pair \tsmall{fa} and \tsmall{la} must be generated with the permutation tuple \tsmall{phi} in order to satisfy Eq.~\eqref{equ:ttt_indices}. Similarly, for $r\leq k<r+s$, stride-based iterators \tsmall{fb}, \tsmall{lb} and \tsmall{fc} are instantiated with the multidimensional input and output iterators \tsmall{B} and \tsmall{C}, respectively. In this case, the permutation tuple \tsmall{psi} is used, in order to generate the correct iterator pair for \tsmall{fb} and \tsmall{lb} with the indices \tsmall{psi[k-r]}. For $k<r+s+q-1$, stride-based iterators are instantiated with the iterators \tsmall{A}, \tsmall{B} and indices \tsmall{phi[k-s]}, \tsmall{psi[k-r]} of the dimensions to be contracted. Finally, when the last branch is reached with $k=r+s+q-1$, the inner product of two fibers of the input arrays are computed. The dimension indices for the instantiation of the stride-based iterators with the multidimensional iterators \tsmall{A} and \tsmall{B} are \tsmall{phi[r+q-1]} and \tsmall{psi[s+q-1]}. 
\begin{example}
Let \tsmall{A}, \tsmall{B} and \tsmall{C} be tensor objects (such as \tsmall{fhg::tensor} or \tsmall{fhg::tensor\_view}) of the same type and layout and let the shape tuples of the tensor objects \tsmall{A}, \tsmall{B} and \tsmall{C} are \tsmall{\{3,4,2\}}, \tsmall{\{5,4,3,6\}} and \tsmall{\{3,1,2\}}, respectively. We can then write the following statement in order to multiply  \tsmall{A} with \tsmall{B}. 
\begin{lstlisting}
unsigned phi[] = {2,0,1}, psi[] = {0,3,1,2};
fhg::ttt(0,3,1, phi, psi, A.mbegin(), B.mbegin(), C.mbegin());
\end{lstlisting}
The first statement initializes  \tsmall{C++} arrays that represents the permutation tuples \tsmall{phi} and \tsmall{psi}. The second statement calls the function \tsmall{fhg::ttt} in Listing~\ref{alg:ttt} and performs the \tsmall{2}-mode tensor-times-tensor multiplication.
\end{example}

Higher-order tensor operations are also provided as member functions of the \tsmall{fhg::tensor} class template and may be combined with pointwise operations. The following code snippet shows two template methods of the \tsmall{fhg::tensor} template class for the tensor-matrix and tensor-vector multiplication. 
\begin{lstlisting}
tensor times_vector(tensor const& rhs,std::size_t mode) const;
tensor times_matrix(tensor const& rhs,std::size_t mode) const;
\end{lstlisting}
The \tsmall{tensor} instance in function \tsmall{times\_matrix} needs to have at most two dimensions. In function \tsmall{times\_vector} one the two extents of \tsmall{tensor} must be one. The second parameter \tsmall{mode} determines the dimension over which the contraction is performed, see equations~\eqref{equ:ttv} and~\eqref{equ:ttm}. Please note that the template class \tsmall{fhg::tensor\_view} provides member functions with the same signature. As we will show with examples, the function signatures for higher-order tensor operations are similar to the function signatures that are provided in~\cite{bader:2006:algorithm862}. The user can therefore easily verify and compare the results of our tensor functions with those provided in~\cite{bader:2006:algorithm862}.
\begin{example}
Consider the following listing, where \tsmall{A}, \tsmall{b}, \tsmall{B} and \tsmall{C} are tensor objects of the same element type and layout. Let the shape tuples of \tsmall{A}, \tsmall{b}, \tsmall{B} and \tsmall{C} be \tsmall{\{3,4,2,6\}}, \tsmall{\{3,1\}}, \tsmall{\{4,5,6\}} and \tsmall{\{2,5\}}, respectively. We can then express the multiplication of the tensor \tsmall{A} with the vector \tsmall{b} along the first dimension and the multiplication of the tensor \tsmall{B} with the matrix \tsmall{C} along the second dimension with the following statement.
\begin{lstlisting}
auto D = A.times_vector(b,1) + B.times_matrix(C,2); //D = ttv(A,b,1)+ttm(A,B,2)
\end{lstlisting}
The resulting temporary array object \tsmall{D} is of type \tsmall{fhg::tensor<float>} and has the shape tuple \tsmall{fhg::shape \{4,2,6\}}. 
\end{example}
The first expression \tsmall{A.times\_vector(b,1)} internally calls the template function \tsmall{ttv} where the mode is zero. The second expression \tsmall{B.times\_matrix(C,2)} calls the template function \tsmall{ttm} which is shown in Listing~\ref{alg:ttm}. Please consider the following member function signature of \tsmall{times\_tensor}.
\begin{lstlisting}
tensor times_tensor(tensor const& rhs,std::size_t q,
                                      std::vector<std::size_t>const& phi, 
                                      std::vector<std::size_t>const& psi)const;
\end{lstlisting}
Function \tsmall{times\_tensor} encapsulates the template function \tsmall{fhg::ttt} presented in Listing~\ref{alg:ttt} and provides a user-friendly interface. The parameters \tsmall{phi} and \tsmall{psi} can have different sizes and determine the permutation of the dimension indices of the input and output arrays. The second parameter \tsmall{q} determines the number of contraction as discussed for the template function \tsmall{fhg::ttt}. 

The possibility to permute the resulting dimension indices of the output array can be regarded as a minor extension of the equivalent functions that are provided in~\cite{bader:2006:algorithm862}. The tensor classes provide two member function that exhibit a simpler interface with the minor restriction of not being able to permute the dimension indices of the output array. 
\begin{lstlisting}
tensor times_tensor(tensor const& rhs,std::vector<std::size_t>const& phi)const;
tensor times_tensor(tensor const& rhs,std::vector<std::size_t>const& phi, 
                                      std::vector<std::size_t>const& psi)const;
\end{lstlisting}
The first function allows to specify one permutation tuple which specifies the permutation for both tensors. In case of the second function, parameter \tsmall{phi} and \tsmall{psi} with equal size specify the permutation of the dimension indices of the left and right hand side tensors. In both cases, the permutation tuples specify the dimension indices to be contracted and can be smaller than the order of the tensors. 

The following listing illustrates a multiplication of two tensors \tsmall{A} and \tsmall{B} with unequal shape and layout tuples where the multiplication is performed according to Eq.~\eqref{equ:ttt}.
\begin{example}
Consider the following listing, where \tsmall{A}, \tsmall{B} and \tsmall{C} are tensor objects of the same type and layout. Let the shape tuples of \tsmall{A}, \tsmall{B} and \tsmall{C} be \tsmall{\{3,4,2,6\}}, \tsmall{\{4,3,2\}} and \tsmall{\{2,3,4\}}, respectively. 
\begin{lstlisting}
auto D = A.times_tensor(B, {2,3}); // C = ttt(A,B,[2:3])
auto E = A.times_tensor(C, {2,3}, {3,1}); // C = ttt(A,B,[2:3],[3,1])
\end{lstlisting}
The first statement multiplies \tsmall{A} with \tsmall{B} with the permutation tuple \tsmall{\{2,3\}} which contracts the dimension \tsmall{2} and \tsmall{3} of the input tensors. The second statement performs a multiplication of the \tsmall{2}-nd, \tsmall{3}-rd and \tsmall{3}-rd, \tsmall{1}-st dimensions of the first and second input tensors, respectively.
\end{example}
The comments denote \tsmall{Matlab} statements that contain user-defined functions of the toolbox discussed in~\cite{bader:2006:algorithm862}. 
%The file \tsmall{example4.cpp} contains both listings where the tensor objects are initialized with a pseudo random generator.
%The tensor template class objects are written into an \tsmall{MATLAB} script which also contains function calls of the \tsmall{MATLAB} tensor toolbox for comparison.
In some cases, a tensor is multiplied with a sequence of vectors and matrices such as in case of the higher-order singular value decomposition~\cite{delathauwer:2000:bestrank1,bader:2006:algorithm862}. The following listing exemplifies the function call where the tensor \tsmall{A} from the previous example is multiplied with multiple vectors.
\begin{example}
Let \tsmall{A} be the tensor from the previous example and let \tsmall{a}, \tsmall{b} and \tsmall{c} be tensor tensor objects of type \tsmall{fhg::tensor<float>} representing vectors. The multiplication of a list of vectors denoted by \tsmall{\{a,b,c\}} with the dimensions given by \tsmall{\{1,2,4\}} is then given by the following statements.
\begin{lstlisting}
fhg::tensor<float> a{3,1}, b{4,1}, c{2,1};
fhg::tensor<float> d = A.times_vectors( {a,b,c}, {1,2,4});
//fhg::tensor<float> d = A.times_vectors( {a,b,c}, 3);
\end{lstlisting}
The commented statement performs the same operation where the last parameter denotes the excluded mode.
\end{example}
Function \tsmall{times\_vectors} internally calls the \tsmall{times\_vector} function starting with the vector \tsmall{c} to the vector \tsmall{a}. The implementation of the higher-order power method in Listing~\ref{alg:hopm} uses the \tsmall{times\_vectors}.
\begin{lstlisting}[float,frame=single,frameround={tttt},numbers=left,numberfirstline=true,caption=\footnotesize Higher-Order Power Method, label=alg:hopm, classoffset=1,morekeywords={Tensor,Value}]
\end{lstlisting}

\begin{comment}
template<class Tensor, class Value>
void hopm(size_t K, Tensor const& A, std::vector<Tensor>& u, std::vector<Value>& l)
{
	for(auto k = 0u; k < K; ++k) {
		for(auto r = 0u; r < A.rank(); ++r){
			u[r]  = A.times_vectors(u, r+1);
			l[r]  = u[r].norm();
			u[r] /= l[r];
		}
	}
}	
\end{comment}
The method is discussed in~\cite{delathauwer:2000:bestrank1} and can be regarded as generalization of the best rank-one approximation for matrices~\cite{bader:2006:algorithm862}. The method estimates the best rank-one approximation of a real valued tensor $\mubA$ of order $p$, by finding a rank-$1$ tensor $\mubB$ composed of a scalar $\lambda$ and unit-norm vectors $\mbu_1,\dots,\mbu_p$ with which the least square cost function
\begin{equation}
f(\mubB) = || \mubA - \mubB ||_2 \quad \text{with} \ \mubB = \lambda \ \mbu_1 \circ \cdots \circ \mbu_p
\end{equation}
is minimized. The first parameter \tsmall{K} is maximum number of iterations of the outer loop. We could additionally insert a convergence criteria such as difference of the previous and current value of $lambda$ as stated in~\cite{delathauwer:2000:bestrank1}. The second parameter is the tensor of type \tsmall{fhg::tensor<T>} that shall be approximated. The third input parameter \tsmall{u} are starting values of the unit-norm vectors. In~\cite{delathauwer:2000:bestrank1} the starting values are set as the most dominant left singular vectors of the matrix unfolding of $\mubA$. The last input parameter \tsmall{l} denote scaling factors $\lambda$.

\section{Conclusions}
\label{sec:conclusion}

We have presented a flexible \tsmall{C++} tensor framework that allows users to conveniently implement tensor algorithms with \tsmall{C++} and to easily verify numerical results with the toolbox presented in~\cite{bader:2006:algorithm862}. The framework contains a multi-layered software stack and applies different types of abstractions at each layer. 

Users can choose the high-level layers of the framework to instantiate tensor classes with arbitrary non-hierarchical data layout, order and dimension extents. Member functions of the tensor classes help to generate views and access multidimensional data supporting all types of non-hierarchical data layouts including the first- and last-order storage formats. The transparency of the data access functions with respect to the data layout can be identified as a unique feature of our framework. 

The lower layers of the software stack provide tensor template functions that are implemented only in terms of multidimensional iterator types separating algorithms and data structures following the design principle of the Standard Template Library. Users of our framework are able to include their own tensor types and extend the functionality of our library without modification of the tensor template classes. We have presented our own multidimensional and stride-based iterator classes and exemplified their usage together with implementations of first- and higher-level tensor operations. 

Member functions of the tensor classes encapsulate tensor template functions and allow to program tensor algorithms in a convenient fashion. We have exemplified their usage and implemented a method for the best rank-one approximation of real valued tensors that has been described in~\cite{delathauwer:2000:bestrank1}. All of the tensor template functions have been implemented in a recursive fashion and execute in-place with no restriction of the contraction mode, order or dimension extents. 

In future, we plan to provide and incorporate in-place tensor functions into our framework for a faster computation. We would like to offer or integrate parallelized and cache-efficient versions of the template functions for different data layouts and investigate their runtime behavior.

\appendix
\section{Software}
\label{app:software}

\subsection{Organization and Usage}
TLib is a header-only library that only depends on the \tsmall{C++} standard library. The main folder of our framework has two subdirectories. The \tsmall{tensor/code} directory contains all library code contents with class declarations and implementations. The \tsmall{tensor/examples} directory contains all examples which demonstrate the instantiation and usage of the tensor class template and tensor functions. The declaration of the class templates and function templates are found in the \tsmall{tensor/code/inc} folder where as the definitions are placed in the folder \tsmall{tensor/code/src}. When working with our tensor library, the inclusion of the \tsmall{code/inc/tensor.h} is sufficient. However, one could also use the tensor function templates without our tensor classes but a different multidimensional array. In such a case, only the corresponding header files can be included. 

\subsection{Examples}
TLib is implemented in \tsmall{C++} using features of the \tsmall{C++14} standard. Users of our framework need to compile the examples in the \tsmall{tensor/examples} with a \tsmall{C++} compiler that supports the \tsmall{C++14} standard. We have tested our framework using \tsmall{gcc} version 6.3.0 and \tsmall{clang} version 4.0.0 compilers. We have provided a \tsmall{Makefile} that generates all executables in the \tsmall{tensor/examples/build} directory. Examples 7 to 11 also generate \tsmall{MATLAB} script files for verification that are placed in the \tsmall{tensor/examples/out} directory. The script files can be directly executed in \tsmall{MATLAB}. Examples 8 to 11 also require the \tsmall{MATLAB} toolbox described in~\cite{bader:2006:algorithm862} with which the numerical results are validated.

% Acknowledgments
\begin{acks}
We are grateful to Fritz Mayer-Lindenberg, Thomas Perschke and Konrad Moren for their helpful comments and suggestions. The authors wish to express special thanks to Volker Schatz for his assistance with various aspects of the project. 
%We are also grateful to the Associate Editor and three anonymous referees for useful criticism and suggestions which helped us to materially improve the manuscript.
\end{acks}

%%%%%%%%%%%%%%%%%%%%%%%%%%%%%%%%%%%%%%%%%%%

\bibliographystyle{ACM-Reference-Format}
\bibliography{literature}
%\newpage
%\printbibliography % biblatex

%\bibliographystyle{plain} % 
%\bibliography{literature}
%\nocite{*}

\end{document}